\documentclass{aa}  
\usepackage{color}
\usepackage{multirow}
\usepackage{txfonts}
\usepackage{animate}
\usepackage[colorlinks, citecolor=blue,urlcolor=blue, linkcolor=blue]{hyperref} 
\begin{document} 
   \title{Probing accretion of ambient cloud material\\ into the Taurus B211/B213 filament}
   \subtitle{}
   \author{
    Y. Shimajiri\inst{1},
    Ph. Andr$\acute{\rm e}$\inst{1},
    P. Palmeirim \inst{2}, 
    D. Arzoumanian  \inst{3}, 
    A. Bracco \inst{4}, 
    V. K$\ddot{\rm o}$nyves \inst{1}, 
    E. Ntormousi \inst{5}, and
    B. Ladjelate\inst{6} 
}

   \institute{
\inst{1}Laboratoire AIM, CEA/DSM-CNRS-Universit$\acute{\rm e}$ Paris Diderot, IRFU/Service d'Astrophysique, CEA Saclay, F-91191 Gif-sur-Yvette, France \email{Yoshito.Shimajiri@cea.fr}\\
\inst{2}Instituto de Astrof\'isica e Ci{\^e}ncias do Espa\c{c}o, Universidade do Porto, CAUP, Rua 
das Estrelas, PT4150-762 Porto, Portugal\\
\inst{3}Department of Physics, Graduate School of Science, Nagoya University Nagoya 464-8602, Japan\\
\inst{4}Nordita, KTH Royal Institute of Technology and Stockholm University, Roslagstullsbacken 23, 10691 Stockholm, Sweden\\
\inst{5}Foundation for Research and Technology (FORTH), Nikolaou Plastira 100, Vassilika Vouton GR - 711 10, Heraklion, Crete, Greece\\
\inst{6}Instituto Radioastronomia Milim$\acute{\rm e}$trica, Av. Divina Pastora 7, Nucleo Central, 18012 Granada, Spain
             }

   \date{Received \today; accepted }

 
  \abstract
   {$Herschel$ observations have emphasized the role of molecular filaments in star formation. 
   However, the origin and evolution of these filaments are not yet well understood, partly because of the lack of kinematic information.}
   {We aim to confirm that Taurus B211/B213 filament is accreting background cloud material from a kinematic viewpoint and to investigate the potential influence of large-scale external effects on the formation of the filament. }
   {To examine whether the B211/B213 filament is accreting background gas due to its gravitational potential, we produced a toy accretion model and compared its predictions to the velocity patterns observed in $^{12}$CO (1--0) and $^{13}$CO (1--0). We also examined the spatial distributions of H${\alpha}$, $Planck$ 857 GHz dust continuum, and HI emission to search for evidence of large-scale external effects.}
   {We estimated the depth of the Taurus cloud around the B211/B213 filament to be $\sim$0.3--0.7 pc under the assumption that the density of the gas is the same as the critical density of $^{13}$CO (1--0). Compared to a linear extent of > 10 pc in the plane of the sky, this suggests that the 3D morphology of the cloud surrounding the B211/B213 filament is sheet-like. Position-velocity ($PV$) diagrams observed in $^{12}$CO (1--0) and $^{13}$CO (1--0) perpendicular to the filament axis show that the emission from the gas surrounding B211/B213 is redshifted to the northeast of the filament and blueshifted to the southwest, respectively, and 
that the velocities of both components approach the velocity of the B211/B213 filament as the line of sight approaches the crest of the filament. 
The $PV$ diagrams predicted by our accretion model are in good agreement with the observed $^{12}$CO (1--0) and $^{13}$CO (1--0) $PV$ diagrams, supporting the scenario of mass accretion into the filament proposed by Palmeirim et al. 
Moreover, inspection of the spatial distribution of the H$\alpha$ and 
$Planck$ 857 GHz emission in the Taurus-California-Perseus region on scales up 
to >200 pc suggests that the B211/B213 filament \textcolor{black}{may have formed} as a result of an expanding supershell generated by the Per OB2 association. 
}
   {Based on  these results, we propose a scenario in which the B211/B213 filament was initially formed by large-scale compression of HI gas and then is now growing in mass due to the gravitational accretion of ambient cloud molecular gas.}


   \titlerunning{Accretion into the Taurus filament}
   \authorrunning{Shimajiri et al.}
   \maketitle
%

\section{Introduction}\label{Sect1}

\begin{figure*}
\centering
\includegraphics[width=170mm, angle=0]{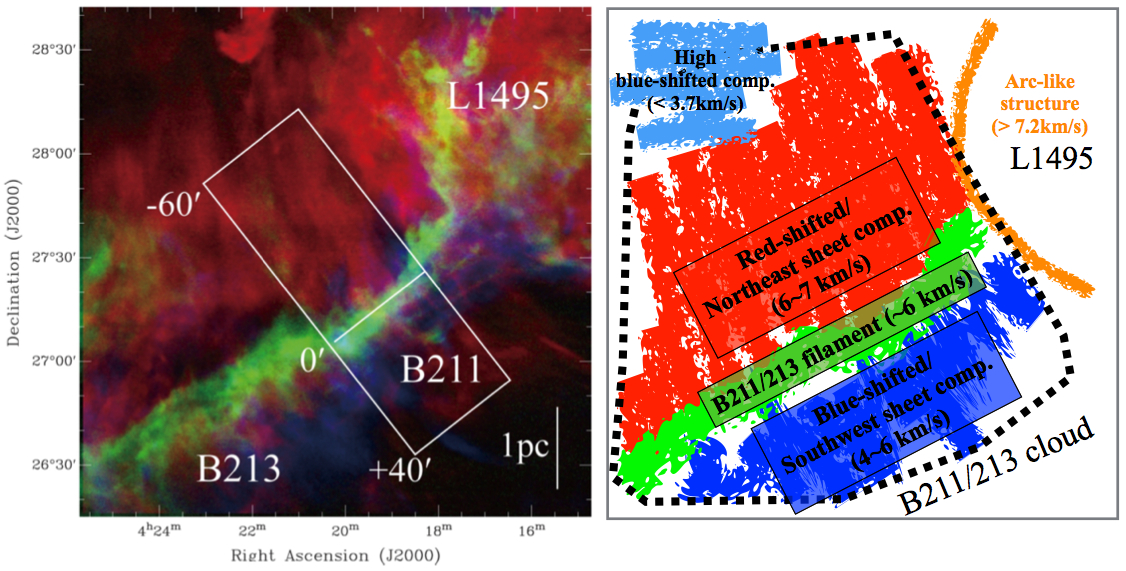}
\caption{($left$)$^{12}$CO (1--0) and $^{13}$CO (1--0) emission observed toward the B211/B213 filament and ($right$) schematic picture of the velocity components. The $^{12}$CO and $^{13}$CO data are from \citet{Goldsmith08}. The panel ($left$) is adopted from \citet{Palmeirim13}.
In panel ($left$), the red color shows the distribution of the $^{12}$CO (1--0) emission with a velocity range of 6.6--7.4 km s$^{-1}$, the green color shows $^{13}$CO (1--0) emission with a velocity range of 5.6--6.4 km s$^{-1}$, and the blue color shows $^{12}$CO (1--0) emission with a velocity range of 4.2--5.5 km s$^{-1}$. The white box perpendicular to the filament axis shows the cut line for the position velocity diagrams shown in Fig. \ref{fig:pv}.}
\label{fig1}
\label{fig:3color}
\end{figure*}

\begin{figure}[t]
\centering
\includegraphics[width=90mm, angle=0]{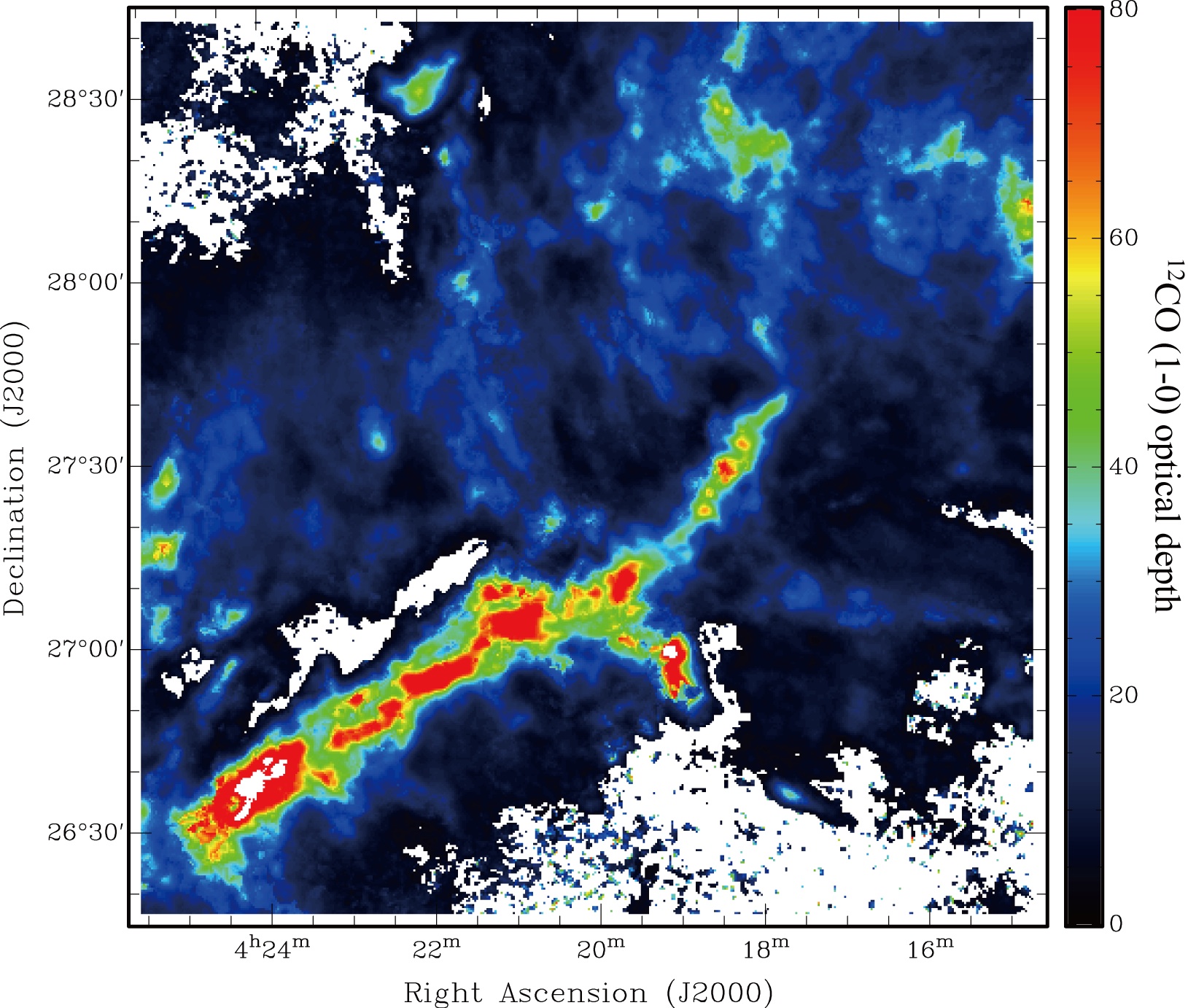}
\caption{Map of $^{12}$CO (1--0) optical depth derived from the \citet{Goldsmith08} $^{12}$CO (1--0) and $^{13}$CO (1--0) data.}
\label{fig:tau}
\end{figure}

The observations of the {\it Herschel} Gould Belt survey (HGBS) have revealed an omnipresence of parsec-scale filaments in molecular clouds and emphasized their importance for solar-type star formation \citep[e.g. ][]{Andre10, Menshchikov10,Arzoumanian11,Palmeirim13}.  
In particular, most $Herschel$ prestellar cores are found to lie in dense (thermally supercritical) filaments, suggesting that cores generally form by filament fragmentation \citep[eg. ][]{Konyves15,Marsh16,Benedettini18}.
Molecular line observations of the velocity field around cores and filaments further support this view \citep{Tafalla15}.
Based on the HGBS results, \citet{Andre14} proposed a filament paradigm for star formation, whereby large-scale compression of interstellar material in supersonic flows generates a quasi-universal web of $\sim$0.1-pc wide filaments in the cold interstellar medium (ISM) and then the denser filaments fragment into prestellar cores by gravitational instability.
Recently, \citet{Shimajiri17} found that the star formation efficiency in dense molecular gas ($A_{\rm v}$ > 8), where filamentary structures dominate the mass budget, is remarkably uniform over a wide range of scales from 1-10 pc to >10 kpc \citep[see also,][]{Gao04,Lada10,Lada12,Chen15}.
Furthermore, \citet{Shimajiri17} proposed that this common star formation efficiency in dense gas results from the microphysics of star formation in filaments \citep[see also][]{Andre14}. This result suggests the existence of a universal "star formation law" converting dense molecular gas into stars along filaments. Therefore, unveiling how molecular filaments grow in mass and fragment is crucial to understanding star formation in filaments.

The B211/B213 filament system is located in the Taurus molecular cloud, which is one of the nearest star-forming regions \citep[$d$$\sim$140 pc,][]{Elias78}. Wide-field mapping observations in $^{12}$CO, $^{13}$CO, C$^{18}$O, N$_2$H$^+$, and SO emission revealed a whole network of filamentary structures in the B211/B213 area \citep{Goldsmith08, Hacar13, Panopoulou14,Tafalla15}. 
\citet{Goldsmith08} and \citet{Palmeirim13} found that many low-density striations are elongated parallel to the magnetic field, and that blueshifted and redshifted components in both $^{12}$CO (1--0) and $^{13}$CO (1--0) emission are distributed to the southwest and the northeast of the B211/B213 filament, respectively, as shown in Fig. \ref{fig:3color}. This morphology was suggestive of mass accretion along magnetic field lines into the B211/B213 filament. To quantify mass accretion, \citet{Palmeirim13} assumed cylindrical geometry and used the observed mass per unit length $M_{\rm line}$ to estimate the gravitational acceleration $\phi$($R$) = 2$GM_{\rm line}/R$ of a piece of gas in free-fall toward the filament (where $R$ and $G$ denote radius from filament center and the gravitational constant, respectively). The free-fall velocity of gas initially at rest at a cylindrical radius $R_{\rm init}\sim$2 pc was estimated to reach 1.1 km s$^{-1}$ when the material reached the outer radius $R_{\rm out}\sim$0.4 pc of the B211/B213 filament. This estimation was consistent with the velocity observed in CO, suggesting that the background gas accretes into the B211/B213 filament owing to the gravitational potential of the B211/B213 filament. However, the velocity structure was not investigated in detail. 
Investigation of the velocity structure is crucial to confirm this suggested scenario from the kinematic viewpoint. This is the topic of the present paper.

The paper is organized as follows: in Sect. \ref{Sect2}, we describe the $^{12}$CO (1--0) and $^{13}$CO (1--0) data, as well as complementary H${\alpha}$, 857 GHz, and HI data. In Sect. \ref{Sect3}, we estimate the optical depth of the $^{12}$CO (1--0) line and present the $^{12}$CO (1--0) and $^{13}$CO (1--0) velocity structures observed in the B211/B213 cloud. In Sect. \ref{Sect4}, we discuss the cloud structure, whether the surrounding material accretes onto the B211/B213 filament from the kinematic viewpoint, and whether the filament is formed by large-scale compression. In Sect. \ref{Sect5}, we summarize our results.

\begin{figure*}
\centering
\includegraphics[width=155mm, angle=0]{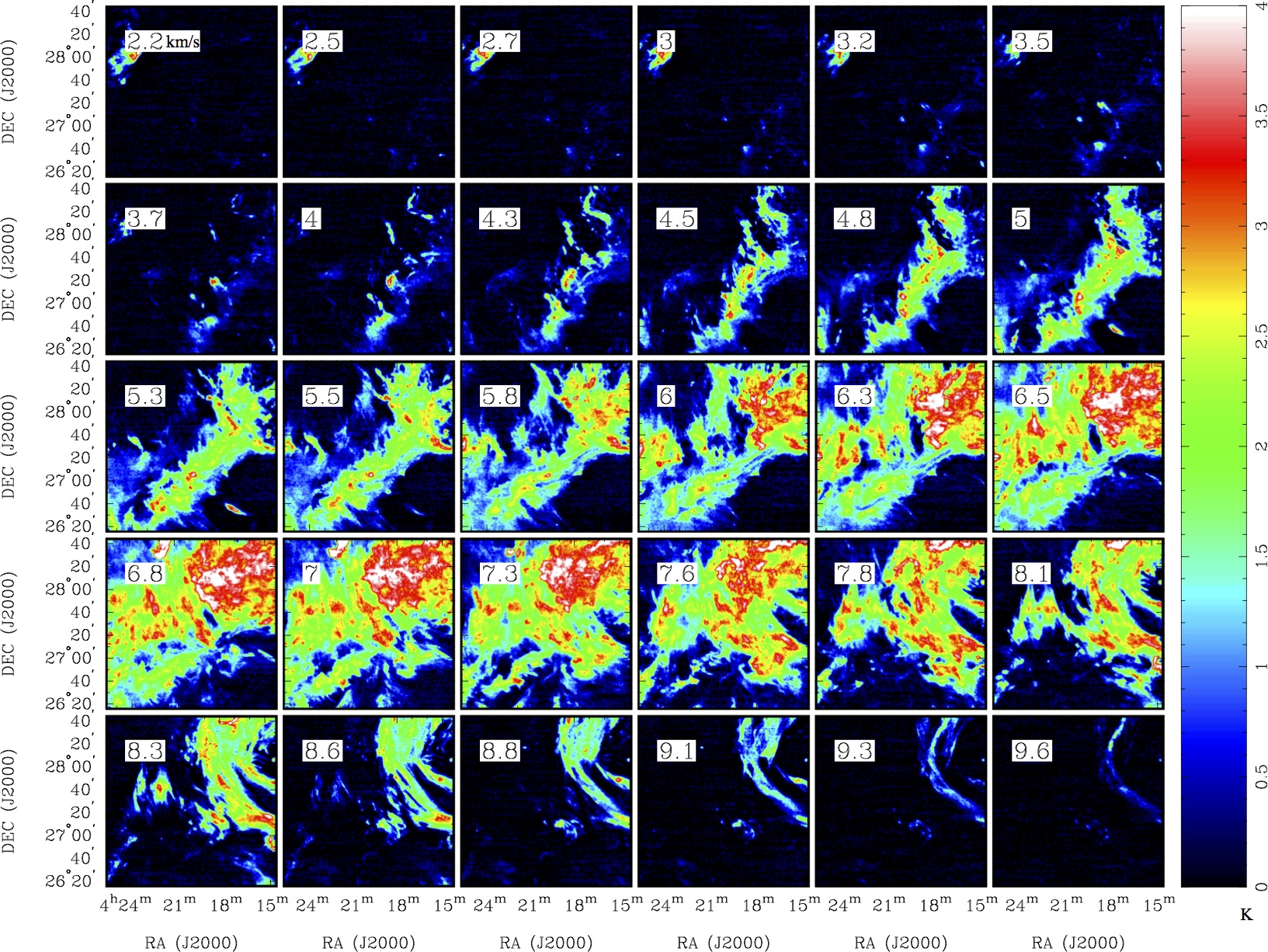}
\includegraphics[width=155mm, angle=0]{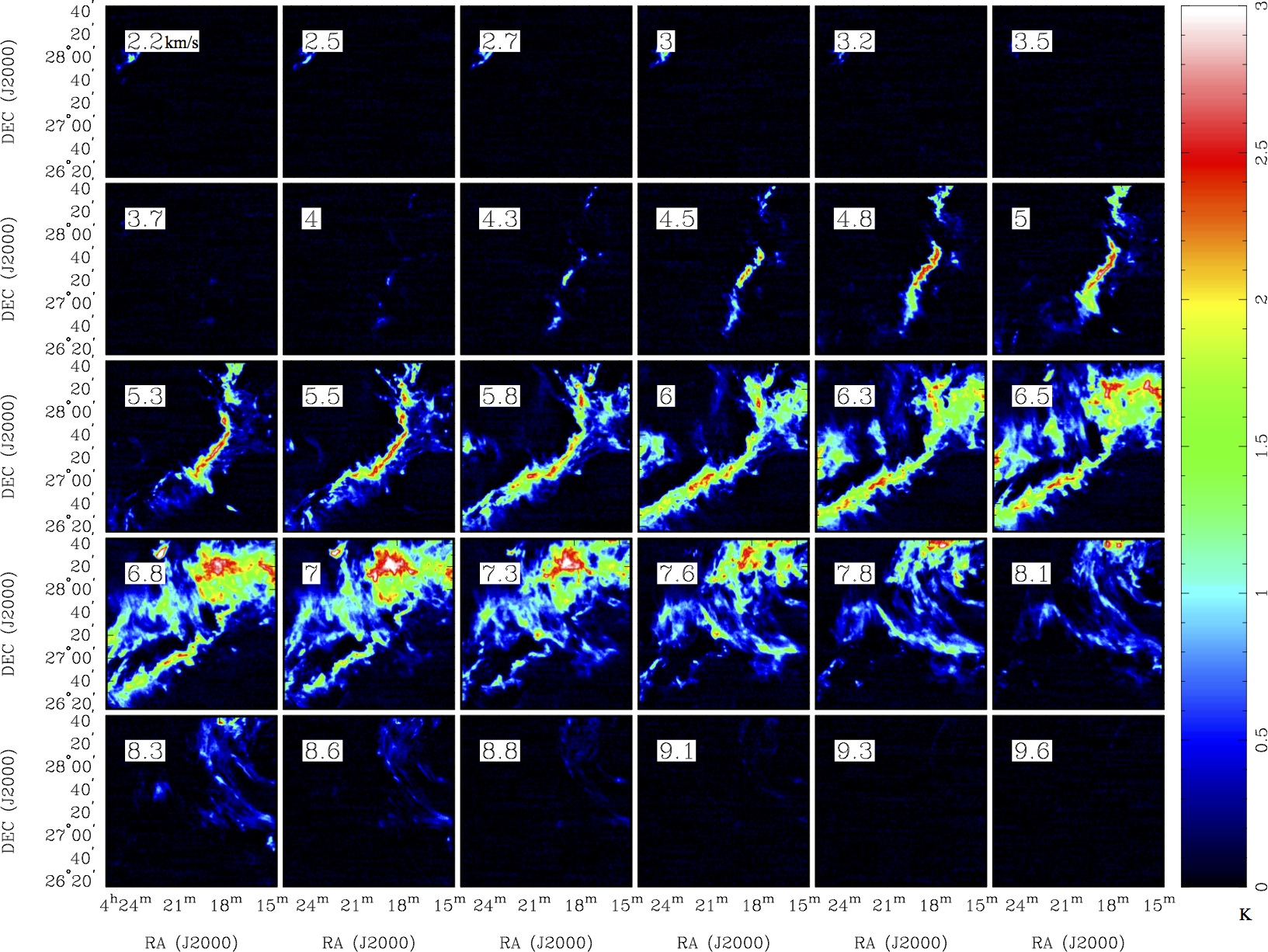}
\caption{Velocity channel maps in the $^{12}$CO ($J$=1--0, $top$) and $^{13}$CO ($J$=1--0, $bottom$) emission lines in units of K obtained from the \citet{Goldsmith08} data. The line of the sight (LSR) velocities (in km s$^{-1}$) are indicated on the top-left corner of each panel. The velocity width of each channel map is 0.3 km s$^{-1}$. }
\label{fig:channel_co}
\end{figure*}

\section{Observational data}\label{Sect2}

In this paper, we used the $^{12}$CO (1--0) and $^{13}$CO (1--0) data obtained by \citet{Goldsmith08, Narayanan08} with the 14 m diameter millimeter-wave telescope of the Five College Radio Astronomy Observatory (FCRAO). The half-power beam  width of the telescope was 45$\arcsec$ for $^{12}$CO (1--0) and 47$\arcsec$ for $^{13}$CO (1--0).
We applied Gaussian spatial smoothing to improve the signal to noise ratio, resulting in an effective beam resolution of $\sim$76$\arcsec$, corresponding to $\sim$0.05 pc at a distance of 140 pc. The velocity resolution of the data is 0.26 km s$^{-1}$ for $^{12}$CO (1--0) and 0.27 km s$^{-1}$ for $^{13}$CO (1--0). The rms noise level is 0.1 K ($T_{\rm A}^*$) for $^{12}$CO (1--0) and 0.05 K ($T_{\rm A}^*$) for $^{13}$CO (1--0), respectively.

As complementary observations of the Taurus cloud region and its large-scale environment, we also used the H${\alpha}$ data\footnote{\url{https://faun.rc.fas.harvard.edu/dfink/skymaps/halpha/data/v1_1/index.html}} of \citet{Finkbeiner03}, as well as $Planck$ 857 GHz\footnote{\url{https://irsa.ipac.caltech.edu/data/Planck/release_1/all-sky-maps/previews/HFI_SkyMap_857_2048_R1.10_survey_2_ZodiCorrected/}} \citep{Planck14} and HI data\footnote{\url{https://www.astro.uni-bonn.de/hisurvey/AllSky_profiles/index.php}} \citep{Kalberla17} from the archive.

\begin{figure*}
\centering
\includegraphics[width=190mm, angle=0]{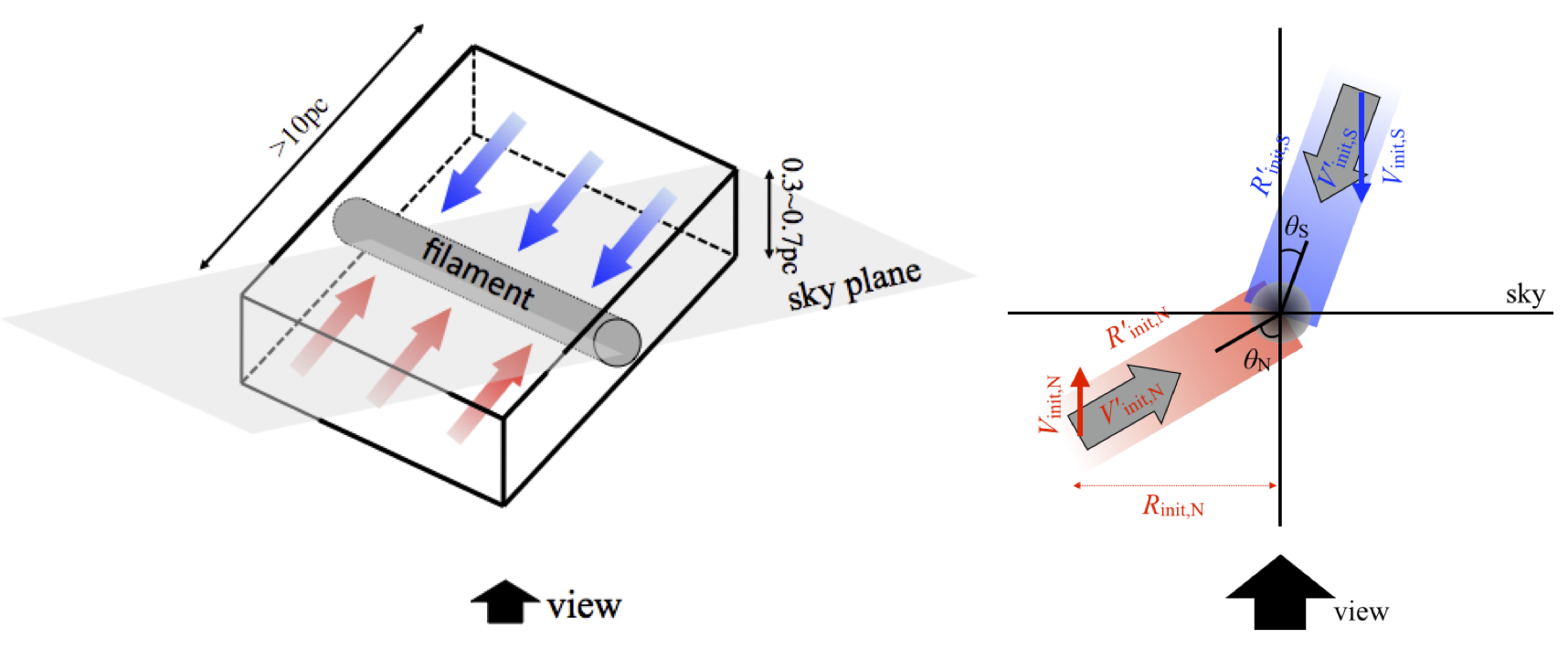}
\caption{($left$) Schematic picture of the structure of the B211/B213 cloud (see Sect. \ref{sect:cloud}), and ($right$) schematic picture describing our toy model of the velocity field (see Sect. \ref{section:model}).}
\label{fig:cloud_structure}
\label{fig:model}
\end{figure*}

\section{Analysis and results}\label{Sect3}

\subsection{$^{12}$CO (1--0) and $^{13}$CO (1--0) optical depths \label{optical_depth}}

The optical depth of the $^{12}$CO (1--0) line was estimated from the FCRAO $^{12}$CO and $^{13}$CO data. Assuming the same excitation temperature for the $^{12}$CO (1--0) and $^{13}$CO (1--0) lines, an isotopic ratio, $R_{\rm i}$ = 62 for $^{12}$C/$^{13}$C \citep{Langer93}, and the same beam filling factor in both lines, we evaluated the optical depth of $^{12}$CO (1--0) using the following equation:

\begin{equation}
\frac{T({\rm ^{13}CO)}}{T({\rm ^{12}CO})}=\frac{1-e^{-\tau({\rm ^{12}CO})/R_{\rm i}}}{1-e^{-\tau(\rm{^{12}CO})}}.
\end{equation}

\noindent Here, $T(\rm{^{12}CO})$ and $\tau(\rm{^{12}CO})$ denote the peak intensity and the optical depth of $^{12}$CO (1--0), respectively.  
While the $^{12}$CO (1--0) emission preferentially traces the diffuse extended cloud, the $^{13}$CO (1--0) emission traces the central B211/B213 filament (see Fig. \ref{fig:3color}). 
The typical inner width of the filaments observed with {\it Herschel} is $\sim$0.1 pc \citep{Arzoumanian11,Arzoumanian18,Palmeirim13}, which is larger than the 0.05 pc effective beam size of the FCRAO data. Thus, assuming the same beam filling factor in $^{12}$CO (1--0) and $^{13}$CO (1--0) is reasonable. 
Figure \ref{fig:tau} displays the resulting map of $^{12}$CO (1--0) optical depth. The optical depth in this map ranges from $\sim$3 to $\sim$300, showing that the $^{12}$CO (1--0) emission is optically thick. In particular, the $^{12}$CO (1--0) optical depth toward the B211/B213 filament itself ($\tau(\rm ^{12}CO)$$\sim$100) is much larger than that found for the surrounding lower density material ($\tau(\rm ^{12}CO)$$\sim$20).

\subsection{$^{12}$CO (1--0) and $^{13}$CO (1--0) velocity channel maps}\label{section:channel}

Figure \ref{fig:channel_co} shows the velocity channel maps observed in $^{12}$CO (1--0) and $^{13}$CO (1--0). 
In the maps for $V_{\rm LSR}$ < 3.7 km s$^{-1}$, both $^{12}$CO (1--0) and $^{13}$CO (1--0) emission is seen in the northeastern part of the maps (RA, DEC = 4:24:00, 28:15:00). In the channel maps for 4.0 < $V_{\rm LSR}$ < 7.3 km s$^{-1}$, enhanced emission is seen toward the B211/B213 filament in both $^{12}$CO (1--0) and $^{13}$CO (1--0). 
The emission at these velocities is likely to be directly associated with the B211/B213 filament. Furthermore, while the emission at 4 km s$^{-1}$ < $V_{\rm LSR}$ < 6 km s$^{-1}$ is distributed to the southwest of the B211/B213 filament, the emission at 6 km s$^{-1}$ < $V_{\rm LSR}$ < 7 km s$^{-1}$ is distributed to the northeast of the filament.
In the channel maps for $V_{\rm LSR}$ > 7.3 km s$^{-1}$, the distribution of the $^{12}$CO (1--0) and $^{13}$CO (1--0) emission is suggestive of an arc-like structure around L1495. 
Figure \ref{fig:3color} ($right$) is a sketch showing the location of each velocity component. 

\section{Modeling of the data and discussion}\label{Sect4}

\subsection{3D morphology of the B211/B213 ambient cloud}\label{sect:cloud}

Here, we discuss the 3D morphology of the material surrounding the B211/B213 filament by comparing the extent of the gas in the plane of the sky and its depth along the line of sight. Hereafter, we refer to the system consisting of the B211/B213 filament and its surrounding gas as the B211/B213 cloud (i.e. red, green, and dark blue areas in Fig. \ref{fig:3color} ($right$)). 

The projected extent of the B211/B213 cloud in the plane of the sky is more than $\sim$10 pc. Taking the viewing angle into account, the real extent of the cloud may be larger. At the same time, we can estimate the depth of the cloud along the line of sight under the assumption that the surrounding material is filled by gas with density exceeding the critical density of the $^{13}$CO (1--0) line, since $^{13}$CO (1--0) emission is observed over the entire mapped area. The critical density of $^{13}$CO (1--0), $n_{\rm critical}^{\rm ^{13}CO}$, may be estimated as follows:

\begin{equation}
n_{\rm critical}^{\rm ^{13}CO} = \frac{A_{\rm ul}}{\sigma_{\rm cross} \nu} = \frac{A_{\rm ul}}{10^{-15}{\rm cm}^{-2} \times 10^4 \sqrt{T_{\rm ex}}}, 
\end{equation}

\noindent where $A_{\rm ul}$, $\sigma_{\rm cross}$, $\nu$, and $T_{\rm ex}$ are the Einstein spontaneous emission coefficient, collision cross section, collision velocity, and line excitation temperature. The values of $A_{10}$ and $\sigma_{\rm cross}$ in the LAMDA database\footnote{\url{http://home.strw.leidenuniv.nl/~moldata/datafiles/13co.dat}} are 6.294$\times$10$^{-8}$ s$^{-1}$ and 10$^{-15}$ cm$^{-2}$. The collision velocity can be calculated as $v$=$\sqrt{3k_{\rm B}T_{\rm ex}/m}$ = $10^{4}\sqrt{T_{\rm ex}}$ cm s$^{-1}$, where $k_{\rm B}$ is the Boltzmann constant and $m$ is hydrogen molecular mass. This leads to a value of 1.7 $\times$ 10$^3$ cm$^{-3}$ for the critical density of $^{13}$CO (1--0) assuming $T_{\rm ex}$ $\simeq$ 14 K.
Here, we assumed that the excitation temperature $T_{\rm ex}$ of the $^{13}$CO (1--0) line is the same as the dust temperature $T_{\rm dust}$ $\sim$ 14K derived by \citet{Palmeirim13} from ${\it Herschel}$ data in the ambient cloud around B211/B213 (red and dark blue area in Fig. \ref{fig1} ($right$)). \citet{Palmeirim13} also estimated the mean {\it Herschel} column density in the material surrounding the B211/B213 filament to be $N_{{\rm H}_2}$ $\simeq$ 1.4 $\times$ 10$^{21}$ cm$^{-2}$. Thus, the depth of the cloud (=$N_{{\rm H}_2}/n_{\rm critical}^{\rm ^{13}CO}$) is estimated to be 0.3 pc.

Recently, \citet{Qian15} independently estimated the depth of the whole Taurus molecular cloud and found a value of $\sim$0.7 pc using the core velocity dispersion (CVD) method. 
With a projected extent of more than 10 pc and a depth of $\sim$0.3 -- 0.7 pc, we conclude that the 3D morphology of the cloud resembles a sheet-like structure (see Fig. \ref{fig:cloud_structure}).
\textcolor{black}{From HC$_3$N (2--1) and (10--9) observations, \citet{Li12} found that the depth of the dense ($\sim$10$^4$ cm$^{-3}$) portion of the B213 region (i.e. the dense 
filament) was $\sim$0.12 pc. This is smaller than our estimate for the depth of the ambient molecular gas, but consistent with the view that the dense inner part of the B213 filament is a cylinder-like structure of $\sim$0.1 pc diameter \citep{Palmeirim13}, embedded in a lower-density sheet-like cloud.
}

\begin{figure}
\centering
\includegraphics[width=80mm, angle=0]{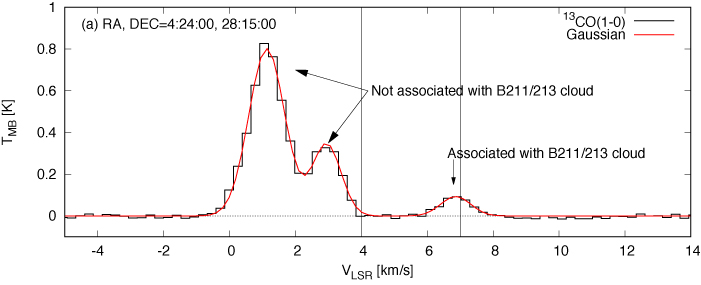}
\includegraphics[width=80mm, angle=0]{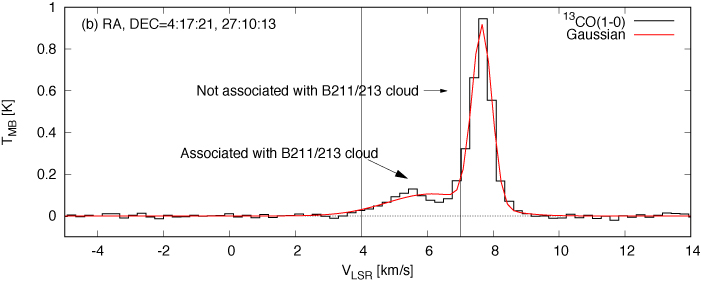}
\caption{Schematic picture of the definition of velocity components associated with the B211/B213 filament. The spectrum in each panel is the $^{13}$CO (1-0) spectrum averaged over a 15$\arcmin$ $\times$ 15$\arcmin$ area with a center position indicated in the top-left corner. The velocity components with a velocity of < 4.0 km s$^{-1}$ or > 7.0 km s$^{-1}$ are regarded as components not associated with the B211/B213 filament. These components are subtracted from the data cube.}
\label{fig:subt}
\end{figure}

\begin{figure*}
\centering
\includegraphics[width=190mm, angle=0]{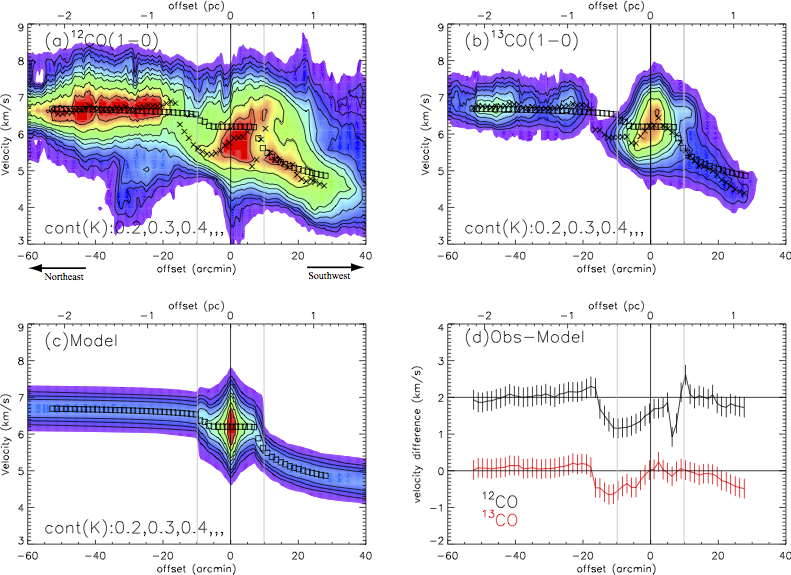}
\caption{Position-Velocity diagram of ($a$) $^{12}$CO (1--0), ($b$) $^{13}$CO (1--0), and ($c$) model and ($d$) velocity offsets between $^{12}$CO (1--0) and $^{13}$CO (1--0) observations and model. Assumed parameters for the accretion model are summarized in Table \ref{Table1}. The cut line of the $PV$ diagrams is indicated in Fig \ref{fig:3color}.
In panels ($a$)-($c$), black squares indicate the peak velocity positions at each offset. In panels ($a$) and ($b$), black crosses are the peak velocity positions at each offset in the model.  
In panel ($d$), lines indicate the velocity offset (black) between $^{12}$CO(1-0) and model and (red) between $^{13}$CO(1-0) and model.  
In panels ($a$)-($d$), black and grey vertical lines indicate offset = 0$\arcmin$ and |offset| < $R_{\rm out}$.}
\label{fig:pv}
\end{figure*}

\subsection{Accretion of background gas into the B211/B213 filament}\label{sect:accretion}

Here, we compare the velocity pattern seen in $^{12}$CO (1--0) and $^{13}$CO (1--0) emission with the prediction of an accretion gas model, in order to investigate whether the B211/B213 filament accretes ambient cloud gas from a kinematic viewpoint.

\subsubsection{Observed position-velocity diagrams}\label{section:pv_discription}

As mentioned in Sect. \ref{section:channel}, the highly blueshifted and redshifted components at $V_{\rm LSR}$ < 3.7 km s$^{-1}$ and $V_{\rm LSR}$ > 7.3 km s$^{-1}$ do not seem to be directly connected to the B211/B213 cloud/filament. To focus on the velocity field of the gas associated with the B211/B213 filament, we subtracted these two components as follows. We applied Gaussian fitting with $N$ Gaussian components to each pixel, where $N$=1, 2, 3, 4, or 5. Wherever the signal to noise (S/N) ratio of the residual peak intensity was less than 5, the fit was deemed to be acceptable and the corresponding spectrum was assumed to consist of $N$ Gaussian components. 
Then, if the peak LSR velocity of a Gaussian component was lower than 4.0 km s$^{-1}$ or higher than 7.0 km s$^{-1}$, the component was not considered to be associated with the B211/B213 filament or cloud and was subtracted from the data cube (see also Fig. \ref{fig:subt} and Fig. \ref{fig:fitting}). Figure \ref{fig:channel_subt} displays the $^{12}$CO (1--0) and $^{13}$CO (1--0) velocity channel maps after subtracting these components. Hereafter, we used these subtracted data cubes.

Figure \ref{fig:pv} shows the resulting position-velocity ($PV$) diagrams in $^{12}$CO (1--0) and $^{13}$CO (1--0) along a direction perpendicular to the B211/B213 filament as indicated in Fig. \ref{fig:3color}.
On these $PV$ diagrams, distinct velocity pattern can be recognized in $^{12}$CO (1--0) and $^{13}$CO (1--0) toward the filament (|offset| < 10$\arcmin$ $\sim$ 0.4 pc). This is probably due to differing optical depths in the two lines. As described in Sect. \ref{optical_depth}, the $^{12}$CO (1--0) optical depth toward the filament is $>$ 50 and much larger than the optical depth toward the outskirts of the filament, suggesting that the $^{12}$CO (1--0) emission only traces the surface of the filament. In the outskirts of the B211/B213 filament (|offset| > 10$\arcmin$), the blueshifted emission is distributed to the southwest (offset > 0$\arcmin$) and the redshifted emission is distributed to the northeast (offset < 0$\arcmin$) of the filament. It can be seen that 
the velocities of the blueshifted and redshifted components approach the velocity of the B211/B213 filament as the offset approaches 0 (i.e. the crest of the filament). 
\textcolor{black}{Transverse velocity gradients perpendicular to the major axis of filaments have been also observed toward several dense filaments in the Serpens cloud \citep{Dhabal18} as well as the main filament in the northwestern part of the L1495 subregion \citep{Arzoumanian18a}.}

\subsubsection{Gas accretion model}\label{section:model}

\begin{table}
\caption{\textcolor{black}{Properties of the three modeled cloud components}}             
\label{Table1}      
\centering                         
\begin{tabular}{c c c c}        
\hline\hline                
Component & Parameter &   \\   
\hline                       
\multirow{4}{*}{Filament$^{\star}$}      & $M_{\rm line}$ & 54 $M_{\odot}$ \textcolor{black}{pc$^{-1}$} $^{\dag}$  \\      
                                            & $n_{\rm H_2}^0$  & 4.5$\times$10$^{4}$ cm$^{-3}$ $^{\dag}$  \\   
                                            & $p$ & 2 $^{\dag}$ \\         
                                            & $R_{\rm flat}$ & 0.03 pc $^{\dag}$ \\      
                                            & $R_{\rm out}$ & 0.4 pc $^{\dag}$ \\  
                                            & $\mathcal{V}_{\rm filament}$ & 6.2 km s$^{-1}$ $^{\ddag}$ \\
\hline \hline
\multirow{3}{*}{Northeastern sheet$^{\star}$} & $\mathcal{V}_{\rm init,N}$ & 6.8 km s$^{-1}$ $^{\ddag}$ \\
                                            & $R_{\rm init,N}'$ &  10 pc \\
                                            & $\theta_{\rm N}$ & 70 deg \\  
\hline \hline
\multirow{3}{*}{Southwestern sheet$^{\star}$} & $\mathcal{V}_{\rm init,S}$ & 4.4 km s$^{-1}$ $^{\ddag}$  \\   
                                            & $R_{\rm init,S}'$ &  10 pc  \\   
                                            & $\theta_{\rm S}$ & 20 deg  \\   
\hline
\hline
\end{tabular}
\tablefoot{(\dag)Adopted from \citet{Palmeirim13}. (\ddag) Peak velocity at $R_{\rm init}$. \textcolor{black}{($\star$) The total masses of the filament, northeastern sheet, and southwestern sheet components are estimated to be $\sim$400 $M_{\odot}$, $\sim$700 $M_{\odot}$, and $\sim$300 $M_{\odot}$, respectively, from the $Herschel$ H$_2$ column density map. }}
\end{table}

The $PV$ diagrams in Fig. \ref{fig:pv} show an asymmetric velocity distribution 
on either side of the 0 position (filament crest), suggesting that the sheet-like ambient cloud surrounding the B211/B213 filament has a different inclination to the plane of the sky to the northeast and the southwest of the filament. 
To investigate whether the B211/B213 filament accretes gas from the ambient cloud, we thus constructed a 3-component toy model (one filament component and two components for the northeastern and southwestern sheets) under the assumption that the sheet components to the northeast (red-shifted) and the southwest (blues-shifted) lie on the near and far sides of the B211/B213 filament, respectively, as shown in Fig. \ref{fig:model}. Our modeling procedure is summarized in the schematic picture shown in Fig. \ref{fig:modeling_flow}.

\subsubsection*{$\bullet$ Model for the central filament component}
First, we produced a model for the filament. 
{\it Herschel} observations of nearby clouds have shown that the radial column density profiles of molecular filaments in the radial direction $R'$ (i.e. perpendicular to the filament crest) can be well described by the following "Plummer-like" function \citep{Arzoumanian11,Palmeirim13}:

\begin{eqnarray}
\begin{tiny}
\begin{split}
\Sigma_p(R')/\mu m_{\rm H} & = \frac{N_{\rm H_2}^0}{[1+(R'/R_{\rm flat})^2]^\frac{p-1}{2}}
&\to 
\rho_{p}(R') = \frac{\rho_{\rm c}}{[1+(R'/R_{\rm flat})^2]^\frac{p}{2}} ,
\end{split}
\end{tiny}
\end{eqnarray}

\noindent where $\rho_{\rm c}$, $\Sigma_p$, $\mu$, $m_{\rm H}$, $N_{\rm H_2}^0$, $p$, and $R_{\rm flat}$ are the central density of the filament, the column density as a function of radius $R'$,  the mean molecular mass, the hydrogen atom mass, the central column density, the index of the power-law density profile at large radii ($R'$ $\gg$ $R_{\rm flat}$), and the radius of the flat inner region, respectively. For the B211/B213 filament, we adopted $N_{\rm H_2}^0$=1.4$\times$10$^{21}$ cm$^{-2}$, $p$=2.0, and $R_{\rm flat}$=0.03 pc from the fitting results of \citet{Palmeirim13}.
We assumed that the filament itself lies in the plane of the sky and that the shape of the intensity profile of the B211/B213 filament as traced in $^{12}$CO (1--0) and $^{13}$CO (1--0) emission is the same as that found in the {\it Herschel} column density map. 
Then, we rescaled the peak integrated intensity to be 2 K km s$^{-1}$ as observed in $^{13}$CO (1--0). 

Approximating the Plummer density profile of the filament by a broken power-law, 
the gravitational potential in the radial direction $R'$ can be expressed as follows\footnote{Here, $R'$ denotes the radius corrected for inclination to the line of sight, where the relation between the corrected radius $R'$ and radius in the sky plane $R$ is $R' = R/\sin(\theta_{\rm N/S})$ and $\theta_{\rm N/S}$ is the inclination angle of the northeastern/southwestern sheet component to the line of sight.} \citep[cf.][]{Hennebelle13}:

\begin{equation}
\begin{tiny}
\phi(R') = 
\begin{cases}
G \rho_{\rm flat} \pi R'^2 & \text{for $R'$ $\le$ $R_{\rm flat}$} \\
G M_{\rm line,flat} \left[1 + 2 \ln\left(\frac{R'}{R_{\rm flat}}\right) + 2 \left(\ln\frac{R'}{R_{\rm flat}}\right)^2 \right] & \text{for $R_{\rm flat}$ < $R'$ $\le$ $R_{\rm out}$} \\
G M_{\rm line,flat} \left[1 + 2 \ln \left(\frac{R_{\rm out}}{R_{\rm flat}}\right) + 2 \left(\ln\frac{R_{\rm out}}{R_{\rm flat}} \right)^2 \right] \\ 
\qquad +2G M_{\rm line} \ln \left(\frac{R'}{R_{\rm out}}\right) & \text{for $R_{\rm out}$ < $R'$}
\end{cases}
\end{tiny}
\end{equation}

\noindent where $\rho_{\rm flat}$ and $R_{\rm out}$ are the density of the filament at $R'$ $\le$ $R_{\rm flat}$ and outer radius of the filament, respectively. We adopted $n_{\rm H_2}^0$=$\rho_{\rm flat}$/$\mu m_{\rm H}$ = $4.5 \times 10^4 \ {\rm cm}^{-3}$, $R_{\rm flat}$ = 0.03 pc, and $R_{\rm out}$ = 0.4 pc from \citet{Palmeirim13} as summarized in Table \ref{Table1}. In the above equation, $M_{\rm line,flat}$ and $M_{\rm line}$ represent the inner and total masses per unit length of the filament and are given by:

\begin{eqnarray}
M_{\rm line,flat} = \rho_{\rm flat} \pi R_{\rm flat}^2 
\end{eqnarray}

\begin{eqnarray}
M_{\rm line} = \rho_{\rm flat} \pi R_{\rm flat}^2 \left[1+2 \ln\left( \frac{R_{\rm out}}{R_{\rm flat}}\right)\right]
\end{eqnarray}

\subsubsection*{$\bullet$ Models for the northeastern and southwestern sheet components}

Second, we produced models for the northeastern and southwestern sheet components assuming that the B211/B213 filament accretes the gas of the sheets as a result of its gravitational potential.

\noindent Taking into account the pressure gradient force, conservation of energy for a parcel of unit mass of the ambient cloud falling onto the central filament may be expressed as follows \citep[cf.][]{Smith94,Smith12}:, 

\begin{eqnarray}
\begin{split}
\scriptsize
\frac{1}{2}(\mathcal{V}_{\rm init,N/S}'^0)^2 \mathalpha{+} 
& \phi(R_{\rm init,N/S}') \mathalpha{+} C_{\rm s,eff}^2 \ln(\rho_{\rm init}) \\
&\mathalpha{=}
\frac{1}{2}\mathcal{V}(R')^2 \mathalpha{+} \phi(R') \mathalpha{+} C_{\rm s,eff}^2 \ln(\rho(R')),
\end{split}
\end{eqnarray}

\noindent where $\mathcal{V}(R)$ is the projected velocity and $C_{\rm s,eff}$ is the effective sound speed. 
The projected velocity $\mathcal{V}(R)$ of the gas flow can thus be expressed as follows:

\begin{eqnarray}
\scriptsize
\mathcal{V}(R)\mathalpha{=}\mathcal{V}_{\rm filament} \mathalpha{\pm}  \sqrt{2\left[\frac{1}{2}(\mathcal{V}_{\rm init,N/S}'^0)^2 \mathalpha{+} \phi(R_{\rm init,N/S}')\mathalpha{-}\phi(R') \mathalpha{+} C_{\rm s,eff}^2 \ln \left(\frac{\rho_{\rm init}}{\rho(R')}\right)\right]} \mathalpha{\times} \cos(\theta_{\rm N/S}),
\end{eqnarray}

\noindent where $\mathcal{V}_{\rm filament}$, $\mathcal{V}_{\rm init,N/S}'^0$, $R_{\rm init,N/S}'$, and $\rho_{\rm init}$ are the systemic velocity of the filament, the velocity of the accreting gas at the initial point corrected for inclination, the initial radius of the accreting gas corrected for inclination, and the volume density at the initial point, respectively. 
Here, we define ${\mathcal{V}_{\rm init,N/S}'^0}$ as $(\mathcal{V}_{\rm init,N/S} - \mathcal{V}_{\rm filament} )/\cos(\theta_{\rm N/S})$, where $\mathcal{V}_{\rm init,N/S}$ is the projected velocity of the northeastern/southwestern sheet component at $R_{\rm init,N/S}'$. 
We adopted $C_{\rm s,eff}$ = 0.9 km s$^{-1}$ from $C_{\rm s,eff} \simeq \delta V_{\rm FWHM}(^{12}{\rm CO})/\sqrt{8\ln2}$, where $\delta V_{\rm FWHM}(^{12}{\rm CO})$ is the $^{12}$CO (1-0) line width (=2.1 km s$^{-1}$) observed toward the B211/B213 filament. 
Wherever the value of $C_{\rm s,eff}^2 \ln \left(\frac{\rho_{\rm init}}{\rho(R')}\right)$ was larger than $\frac{1}{2}{(\mathcal{V}_{\rm init,N/S}'^0)^2} + \phi(R_{\rm init,N/S}')-\phi(R')$, we adopted $\mathcal{V}(R) = \mathcal{V}_{\rm filament}$. 
The $Herschel$ observations show that the density profile of the B211/B213 filament is proportional to $R'^{-2}$ at $R'$ $\le$ $R'_{\rm out}$ and has a shallower slope at $R'$ $\ge$ $R'_{\rm out}$ \citep{Palmeirim13}. Furthermore, the slope for the southwestern sheet component is slightly steeper than the slope for the northeastern sheet component. At $R'$ > $R'_{\rm init}$, the gas density in the model was assumed to be constant.

\begin{figure}
\centering
\includegraphics[width=95mm, angle=0]{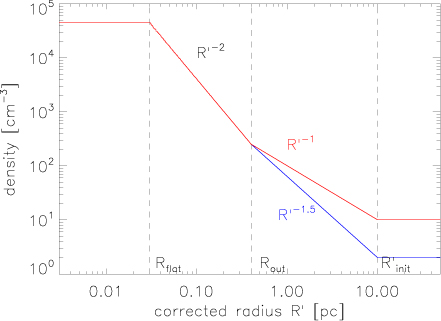}
\caption{Assumed density profile for the 3-component model. Red and blue lines indicate the densities for the northeastern and southwestern sheet components, respectively.}
\label{fig:density}
\end{figure}

To summarize, we assumed the following density distribution as a function of radial direction $R'$ (see Fig. \ref{fig:density}):

For the northeastern sheet component,

\begin{eqnarray}
\footnotesize
\rho(R') = 
\begin{cases}
\rho_{\rm flat} &\text{for $R'$ $\le$ $R_{\rm flat}$} \\
\rho_{\rm flat} \left(\frac{R'}{R_{\rm flat}}\right)^{-2} &\text{for $R_{\rm flat}$ < $R'$ $\le$ $R'_{\rm out}$} \\
\rho_{\rm flat} \left(\frac{R_{\rm out}}{R_{\rm flat}}\right)^{-2}\left(\frac{R'}{R_{\rm out}}\right)^{-1.0} &\text{for $R'_{\rm out}$ < $R'$ $\le$ $R'_{\rm init}$} \\
{\rm constant} = \rho_{\rm flat} \left(\frac{R_{\rm out}}{R_{\rm flat}}\right)^{-2}\left(\frac{R'_{\rm init}}{R_{\rm out}}\right)^{-1.0} & \text{for $R'_{\rm init}$ < $R'$} \\
\end{cases}
\end{eqnarray}

For the southwestern sheet component,

\begin{eqnarray}
\footnotesize
\rho(R') = 
\begin{cases}
\rho_{\rm flat} &\text{for $R'$ $\le$ $R_{\rm flat}$} \\
\rho_{\rm flat} \left(\frac{R'}{R_{\rm flat}}\right)^{-2} &\text{for $R_{\rm flat}$ < $R'$ $\le$ $R'_{\rm out}$} \\
\rho_{\rm flat} \left(\frac{R_{\rm out}}{R_{\rm flat}}\right)^{-2}\left(\frac{R'}{R_{\rm out}}\right)^{-1.5} &\text{for $R'_{\rm out}$ < $R'$ $\le$ $R'_{\rm init}$} \\
{\rm constant} = \rho_{\rm flat} \left(\frac{R_{\rm out}}{R_{\rm flat}}\right)^{-2}\left(\frac{R'_{\rm init}}{R_{\rm out}}\right)^{-1.5} & \text{for $R'_{\rm init}$ < $R'$} \\
\end{cases}
\end{eqnarray}

We also assumed that both sheet components have integrated intensities of $\sim$1 K km s$^{-1}$ as observed in $^{13}$CO (1-0). To get a good agreement between the models and the observations (see Appendix \ref{appendix:inclination}), 
we adopted $\theta_{\rm N}$ = 70$^{\circ}$, $\mathcal{V}_{\rm init,N}$ = 6.8 km s$^{-1}$, and $R_{\rm init,N}'$ = 10 pc for the northeastern sheet component and $\theta_{\rm S}$ = 20$^{\circ}$, $\mathcal{V}_{\rm init,S}$ = 4.4 km s$^{-1}$, and $R_{\rm init,S}'$ = 10 pc for the southwestern sheet component. The parameters of our model are summarized in Table \ref{Table1}. \textcolor{black}{For simplification, we assumed constant inclinations to the line of sight, of 70$^\circ$ for the northeastern sheet component and 20$^\circ$ for the southwestern sheet component, in our model. In reality, however, the inclinations of the two components may vary smoothly with radius from the B211/B213 filament and match on the filament crest.
}

\subsubsection*{$\bullet$ Combined 3-component model}
We first generated an integrated intensity distribution and a peak velocity field for each of the three components with IDL (Interactive Data Language). 
Using the MIRIAD task $velimage$\footnote{$velimage$ makes a data cube $output\_cube(x,y,v_{\rm centroid})$ from an input integrated  intensity image $input\_intensity(x,y)$, input centroid velocity image $input\_velocity(x,y)$, and dispersion $\sigma$. The $v_{\rm centroid}$-values are the centroid velocity and are input as an ($x,y$) image. The output cube image is produced as $output\_cube(x,y,v_{\rm centroid}) = input\_intensity(x,y) \times \exp(-(v_{\rm centroid}-input\_velocity(x,y))^2/(2\sigma^2)))$.}, 
we then produced individual data cube components for the filament and the two sheet components
assuming uniform velocity dispersions of 1.3 km s$^{-1}$ for the filament and 0.9 km s$^{-1}$ for the sheet components. The velocity dispersions were obtained from fitting the observed $^{13}$CO (1--0) spectra. 
\textcolor{black}{One of the reasons for the larger velocity dispersion observed in $^{13}$CO toward the central filament may be that the B211/B213 filament contains several velocity subcomponents \citep[or "fiber-like" structures, ][]{Hacar13}, possibly 
as a result of accretion-driven turbulence \citep[cf. ][]{Hennebelle13,Heitsch13,Andre14}\footnote{In recent numerical simulations of this process, 
\citet{Seifried15} did find that the accretion flow increases the velocity dispersion of the central filament, and \citet{Clarke17} suggested that fiber-like structures could be produced as a result of the vorticity generated by an inhomogeneous accretion flow.}.}

Finally, we used IDL to co-add the three individual data-cube components and produce a combined model data cube.

\subsubsection*{$\bullet$ Large-scale kinematic model}

We adopted initial velocities (corrected for inclination) of {$\mathcal{V}_{\rm init,N}'^0$ = 1.8 km s$^{-1}$ $(=[\mathcal{V}_{\rm init,N} - \mathcal{V}_{\rm filament}]/\cos(\theta_{\rm N}))$} for the northeastern sheet component and {$\mathcal{V}_{\rm init,S}'^0$ = $-$1.9 km s$^{-1}$ $(=[\mathcal{V}_{\rm init,S} - \mathcal{V}_{\rm filament}]/\cos(\theta_{\rm S}))$} for the southwestern sheet component. 
This almost symmetric initial velocity pattern after correction for inclination is suggestive of gravitational accretion.
If the accreting gas comes from far away positions $R_{\rm far,N/S}'$ ($\gg$ $R_{\rm init}'$) and is accelerated by the gravitational potential of the B211/B213 filament/cloud, the line of sight velocity at $R_{\rm far,N/S}'$ is likely to be similar to $\mathcal{V}_{\rm filament}$. The positions $R_{\rm far,N/S}'$ for the northeastern and southwestern sheet components can be estimated from the equation of 
$\mathcal{V}_{\rm init,N/S}'=\sqrt{2[\phi(R_{\rm far,N/S}')-\phi(R_{\rm init,N/S}')]} \times \cos(\theta_{\rm N/S})$ since the pressure density gradient is probably small and can be neglected. We adopted $R_{\rm init,N}'$ = 10 pc, and $R_{\rm init,S}'$ = 10 pc, respectively. Thus, assuming that the initial velocities are entirely generated by gravitational acceleration, the surrounding gas for the northeastern and southwestern sheet components would have to come from $R_{\rm far,N}'$ ($R_{\rm far,N}$) = 270 (260) pc and $R_{\rm far,S}'$ ($R_{\rm far,S}$) = 520 (180) pc (see Fig. \ref{fig:pv_large}). Here, for simplification, we did not include the mass of the sheets when estimating the gravitational potential. Thus, these $R_{\rm far,N/S}'$ values should be considered upper limits. The HI emission observed at $V_{\rm LSR}$ $\sim$ 6 km s$^{-1}$, which corresponds to the systemic velocity of the B211/B213 filament, has an extended emission with an extent of several $\times$ 100 pc which is consistent with the above value of $R_{\rm far,N}'$ (see also Sect. \ref{Sect:compress} and Fig. \ref{fig:channel_hi}). Thus, one of the reasons why the initial velocities at $R_{\rm init,N/S}'$ in the northeastern and southwestern sheet components differ from the velocity of the filament may be the large-scale effect of the gravitational potential of the B211/B213 cloud/filament. We will discuss another possible explanation in Sect. \ref{Sect:compress}.

\subsubsection{Comparing the combined model with the observations}\label{section:comp_model}

\begin{figure}
\centering
\includegraphics[width=80mm, angle=0]{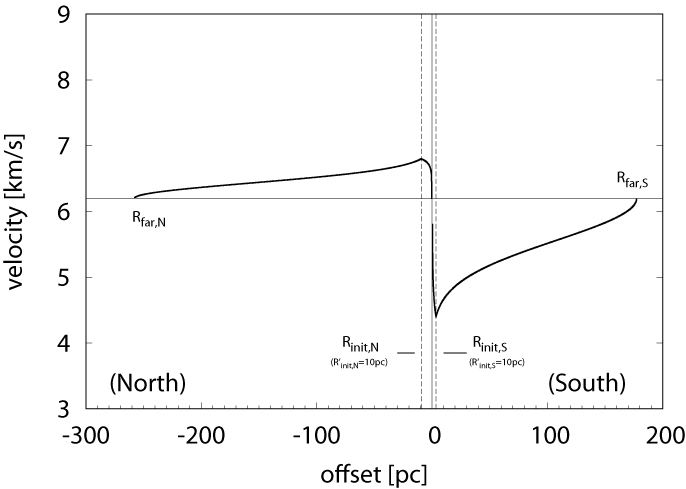}
\caption{Position-velocity diagram of the model for the large scale. $\theta_{\rm N}$=70$^{\circ}$ and $\theta_{\rm S}$=20$^{\circ}$ are assumed. Fig. \ref{fig:pv}  corresponds to -2.4 pc < offset < 1.6 pc in this figure. The vertical dashed lines mark $R_{\rm init,N}$ = -9.4 pc ($R'_{\rm init,N}$ = 10 pc) and $R_{\rm init,S}$ = 3.4 pc ($R'_{\rm init,S}$ = 10 pc).}
\label{fig:pv_large}
\end{figure}

The synthetic $PV$ diagram predicted by the model is shown in Fig. \ref{fig:pv} ($c)$ for comparison with the $PV$ diagrams observed in $^{12}$CO (1--0) and $^{13}$CO(1--0) (Figs. \ref{fig:pv}($a$) and \ref{fig:pv}($b$)). A good quantitative agreement especially with the $^{13}$CO (1--0) diagram can be seen. 
In particular, in both the model and the observed $PV$ diagrams, the velocity of the gas surrounding the B211/B213 filament (red and dark blue area in Fig. \ref{fig:3color}($right$)) approaches the systemic velocity of the B211/B213 filament as the positional offset approaches 0 (i.e. the filament crest). 
While the gas is accelerated by the gravitational potential of the filament/cloud at large scales (several$\times$10 pc), it is decelerated by the pressure gradient force of the dense filament at small scales (several pc) (see Fig. \ref{fig:pv_large} and Fig. \ref{fig:ob2}).
The good agreement between the model and the data indicates that observational kinematic constraints are consistent with the B211/B213 filament accreting background cloud material as a result of its gravitational potential.
This provides strong support to the scenario of mass accretion along magnetic field lines into the filament proposed by \citet{Palmeirim13}. 
The mass accretion rate onto the B211/B213 filament was estimated to be 27-50 $M_{\odot}$ pc$^{-1}$ Myr$^{-1}$ by \citet{Palmeirim13}, suggesting that it took $\sim$1--2 Myr to form the filament.
Thus, accretion of gas from the ambient cloud in B211/B213 likely plays a key role in the evolution of the filament. 

\subsection{Formation of the B211/B213 filament by large-scale compression}\label{Sect:compress}

\begin{figure}
\centering
\includegraphics[width=80mm, angle=0]{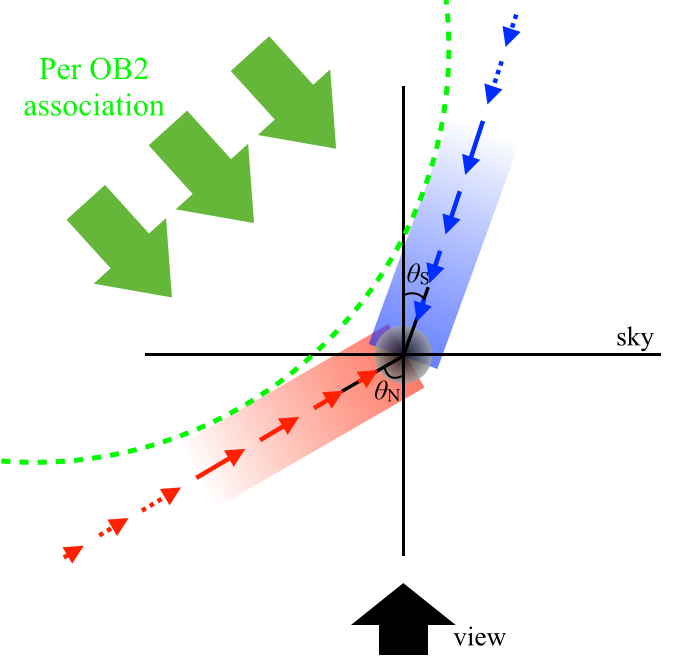}
\caption{
Schematic picture of the relation between the B211/B213 cloud and Per OB2 association. The black arrows indicate the line of sight. The horizontal line indicate the sky plane. Red and blue arrows indicate the direction of the gas accretion in the northeastern and southwestern sheet components, respectively. Green arrows indicate the direction of the compression by Per OB2 association. $\theta_{\rm N}$ and $\theta_{\rm S}$ are the inclination angles of the northeastern and southwestern sheet components to the line of sight. Red and blue arrows of length scaling quantitatively with the magnitude velocity field indicate the direction of the acceleration flow of ambient cloud material. 
}
\label{fig:ob2}
\end{figure}

\begin{figure}
\centering
\includegraphics[width=90mm, angle=0]{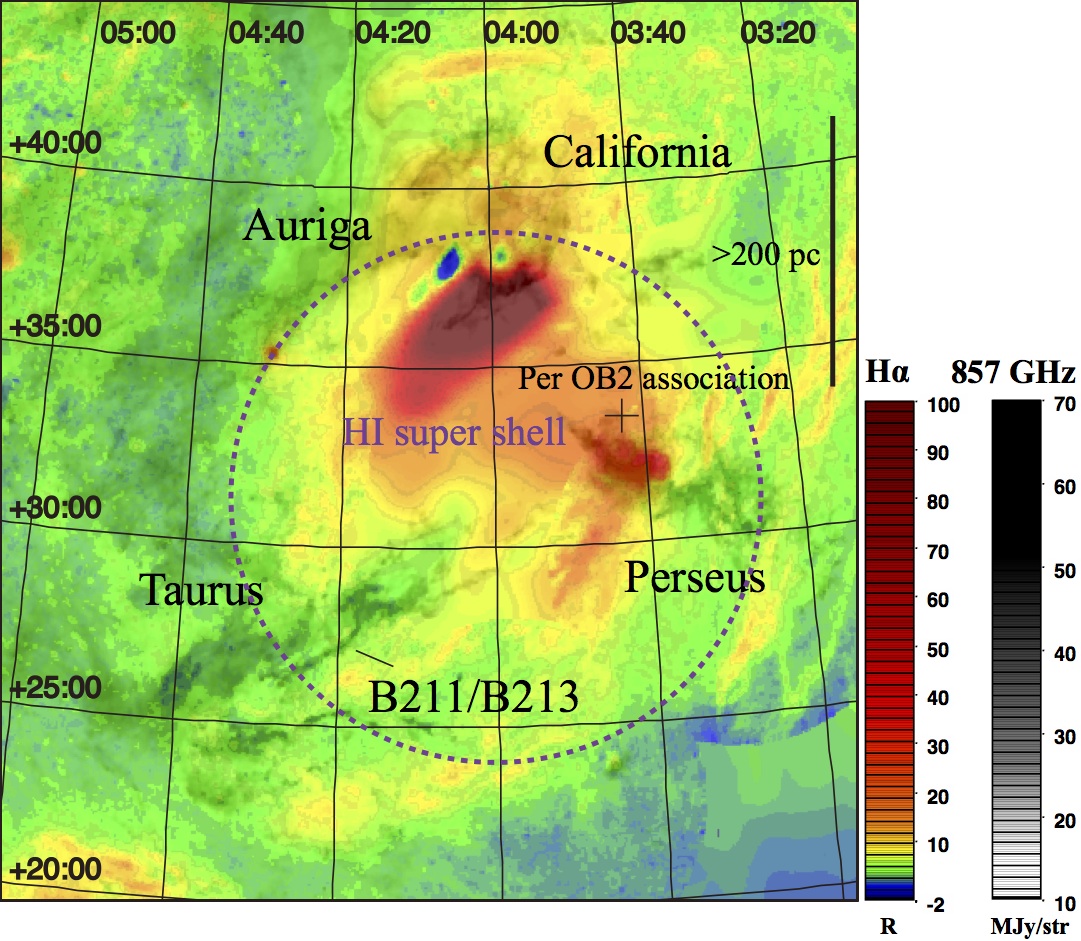}
\caption{Distributions of the H${\alpha}$ \citep[color,][]{Finkbeiner03} and 857 GHz dust \citep[grey,][]{Planck14} emission. The units of the H${\alpha}$ and 857 GHz maps are R (Rayleigh, 4$\pi$ $\times$ 10$^{-4}$ photons cm$^{-2}$ s$^{-1}$ sr$^{-1}$) and MJy str$^{-1}$, respectively.
The magenta dashed circle indicates a HI supershell \citep[][]{Lim13}. The diameter of the HI supershell might be $>$ 200 pc since the distances to the Taurus and Perseus clouds are 140 pc and 340 pc, respectively. The distribution of HI emission is shown in Figs. \ref{fig:channel_hi} and \ref{fig:3color_hi}. See also \ref{fig:halpha_appendix}.}
\label{fig:halpha}
\end{figure}

As described in Sect. \ref{sect:accretion}, we adopted different inclinations for the northeastern sheet component ($i$=70$^{\circ}$) and for the southwestern sheet component ($i$=20$^{\circ}$) in our model to get a good agreement with the observations. This suggests that the B211/B213 cloud is actually shaped like a shell (see Fig. \ref{fig:ob2}). One possibility is that this shell-like structure was produced by large-scale compression. 

In this section, we try to investigate whether the cloud surrounding the B211/B213 filament is affected by large-scale flow phenomena using wide-field H${\alpha}$ maps tracing gas ionized by massive stars \citep{Finkbeiner03}, the $Planck$ 857 GHz dust continuum map tracing cold dust \citep{Planck14}, and HI map tracing lower density atomic gas \citep{HI4PI16}.

Figure \ref{fig:halpha} (see also Figs \ref{fig:channel_hi} and \ref{fig:3color_hi}) compare the spatial distributions of the H${\alpha}$ and 857 GHz emission in the Taurus-California-Perseus region (e.g. Taurus, Auriga, California, and Perseus). The 857 GHz dust emission traces each molecular cloud and exhibits a hole-like structure. This hole-like structure can also be seen in HI emission as shown in Fig. \ref{fig:channel_hi} and Fig. \ref{fig:3color_hi}. 
The H${\alpha}$ emission fills the hole-like structure seen in the 857 GHz dust emission near the center of the field. 
The Taurus, California, and Perseus molecular complexes traced by the 857 GHz dust emission are distributed at the edge of the hole-like structure. \citet{Lim13} also found evidence of a shell-like structure using dust extinction and $^{12}$CO (1--0) maps. 
The hole-like structure may result from the expansion of a large-scale supershell  produced by a supernova in the Per OB2 association that compresses the Taurus cloud from the far side \citep{Olano87, Bally08}. 
An H${\alpha}$ absorption feature is detected toward the Taurus cloud (see Fig. \ref{fig:halpha}  and Fig. \ref{fig:halpha_appendix}), suggesting that the Taurus cloud lies at the front surface of the large-scale supershell produced by the Per OB2 association. 
The distance to the Per OB2 association is estimated to be 340 pc from the Sun \citep{Cernis93}, while the distance to the Taurus cloud is $\sim$140 pc \citep{Elias78}. These distances are consistent with the Taurus cloud lying in front of the Per OB2 association.  
The B211/B213 filament also appears to be in front of the HI shell \citep[see Fig. 10 in][]{Chapman11}. This morphology suggests that the B211/B213 filament may have  formed as a result of an expanding supershell. 
This may provide another reason for the different initial gas velocities differed for the northeastern and southwestern sheet components besides large-scale  acceleration by the gravitational potential of the B211/B213 cloud (see Sect. \ref{section:comp_model}). 
The Local Bubble surrounding the Sun might also compress the Taurus cloud from the opposite direction. 
The Local Bubble surrounding the Sun was produced by supernovae \citep{Snowden98,Sfeir99} and the wall of the Local Bubble is located close to the Taurus cloud \citep{Konyves07,Lallement14}.

Interestingly, the $Planck$ 353 GHz data show variations in the polarization fraction (i.e., polarized intensity/total intensity) across the B211/B213 filament, with lower and higher polarization fractions in the southwestern and northeastern parts of the filament, respectively \citep[see Fig. 10 in ][]{Planck16}. If the gas surrounding the filament is shaped as a shell-like structure with an ordered magnetic field in the plane of each sheet component and if the southwestern sheet component is oriented closer to the line of the sight compared to the northeastern sheet component (cf. Fig. \ref{fig:model}($right$)), the polarization fraction is expected to be lower in the southwestern area (dark blue in Fig. \ref{fig1}($right$)) than in the northeastern area (red in Fig. \ref{fig1}($right$)) assuming uniform dust grain properties. Moreover, both the polarization fraction and the polarization angle show smooth variations across the filament, which is consistent with the northeastern and southwestern sheets being curved (i.e., shell-like). The $Planck$ polarization results are therefore support  the present model.

Using magnetic magnetohydrodynamic (MHD) numerical simulations, \citet{Inutsuka15} and \citet{Inoue18} have argued that multiple compressions associated with expanding bubbles can create star-forming filamentary structures within sheet-like molecular cloud. A similar model of anisotropic filament formation in shock compressed layers has been proposed by \citet{Chen14}, also based on MHD simulations. Such anisotropic filament formation model naturally account for transverse velocity gradients across the B211/B213 filament (see Fig. \ref{fig:3color}) and other dense molecular filaments \citep{Dhabal18}, and are good agreement with the observational picture presented.

Based on these considerations, we propose the following scenario for the formation and evolution of the B211/B213 filamentary system:

\begin{enumerate}

\item A large-scale flow associated with the Per OB2 supershell compressed and deformed the cloud centered on the B211/B213 filament and created a bent shell-like structure.  

\item Owing to its strong gravitational potential, the B211/B213 filament is growing in mass due to accretion of background gas from the surrounding shell-like structure.

\end{enumerate}

\section{Conclusions}\label{Sect5}

To examine whether the B211/B213 filament is accreting gas from the surrounding cloud, we investigated the velocity patterns observed in the $^{12}$CO (1--0) and $^{13}$CO (1--0) lines. Our main findings may be summarized as follows: 

\begin{enumerate}
\item The optical depth of the $^{12}$CO (1--0) line was estimated to be $\sim$3--300. The $^{12}$CO optical depth toward the B211/B213 filament is much larger than that toward the outskirts of the filament. The position-velocity diagrams observed in $^{12}$CO (1--0) and $^{13}$CO (1--0) exhibit different velocity patterns close to the filament, which is likely due to different optical depths.

\item The $^{12}$CO (1--0) and $^{13}$CO (1--0) emission from the B211/B213 filament are seen at an LSR velocity of $\sim$6 km s$^{-1}$. In the northeastern and southwestern parts of the B211/B213 filament, the $^{12}$CO (1--0) and $^{13}$CO (1--0) emission are redshifted and blueshifted, respectively. The line of sight velocities are gradually approaching the systematic velocity of the filament as one gets closer to the filament. 

\item The linear extent of the cloud around the B211/B213 filament is more than 10 pc in the plane of the sky. In contrast, the depth of the cloud along the line of sight is estimated to be $\sim$0.3--0.7 pc (=$N_{\rm H_2}$/$n_{\rm critical}^{\rm ^{13}CO}$) under the assumption that the density of the surrounding material is the same as the critical density of $^{13}$CO (1--0). These results suggest that the 3D morphology of the gas cloud surrounding the B211/B213 filament is sheet-like.

\item To investigate whether the B211/B213 filament is in the process of accreting the surrounding gas material, we compared the velocity patterns observed in $^{12}$CO (1--0) and $^{13}$CO (1--0) with our 3-component model. The predictions of the  model were found to be in good agreement with the distribution of $^{12}$CO (1--0) and $^{13}$CO (1--0) emission in the observed position-velocity diagrams, supporting the scenario of mass accretion along magnetic field lines into the B211/B213 filament proposed by \citet{Palmeirim13}.

\item From an inspection of the wide-field spatial distributions of H${\alpha}$ and 857 GHz dust emission in the Taurus-California-Perseus region, we concluded that the B211/B213 filament was probably formed as a result of the  expansion of a large-scale supershell originated in the Per OB2 association.
This scenario provides a simple explanation for the different inclinations of the northeastern and southwestern sheet components inferred from our modeling analysis.

\item Based on these results, we propose that a) large-scale compression(s) generated by the Per OB2 association initially formed the B211/B213 filament system, and b) accretion of ambient gas material due to the gravitational potential of the filament is now responsible for the growth of the filament.

\end{enumerate}

\begin{acknowledgements}
This work was supported by the ANR-11-BS56-010 project ``STARFICH" and the European Research Council under the European Union's Seventh Framework Programme (ERC Advanced Grant Agreement no. 291294 --  `ORISTARS'). YS also received support from the ANR (project NIKA2SKY, grant agreement ANR-15-CE31-0017). P.~P. acknowledges support from the Funda\c{c}\~ao para a Ci\^encia e a Tecnologia of 
Portugal (FCT) through national funds (UID/FIS/04434/2013) and by FEDER through 
COMPETE2020 (POCI-01-0145-FEDER-007672) and also by the fellowship SFRH/BPD/110176/2015 
funded by FCT (Portugal) and POPH/FSE (EC).
This research has made use of "Aladin sky atlas" developed at CDS, Strasbourg Observatory, France \citep{Bonnarel00, Boch14}. 
\end{acknowledgements}

\bibliographystyle{aa}
\bibliography{B211_accretion.bbl}

\appendix{}
\section{Complementary figures}

\begin{figure*}
\centering
\includegraphics[width=140mm, angle=0]{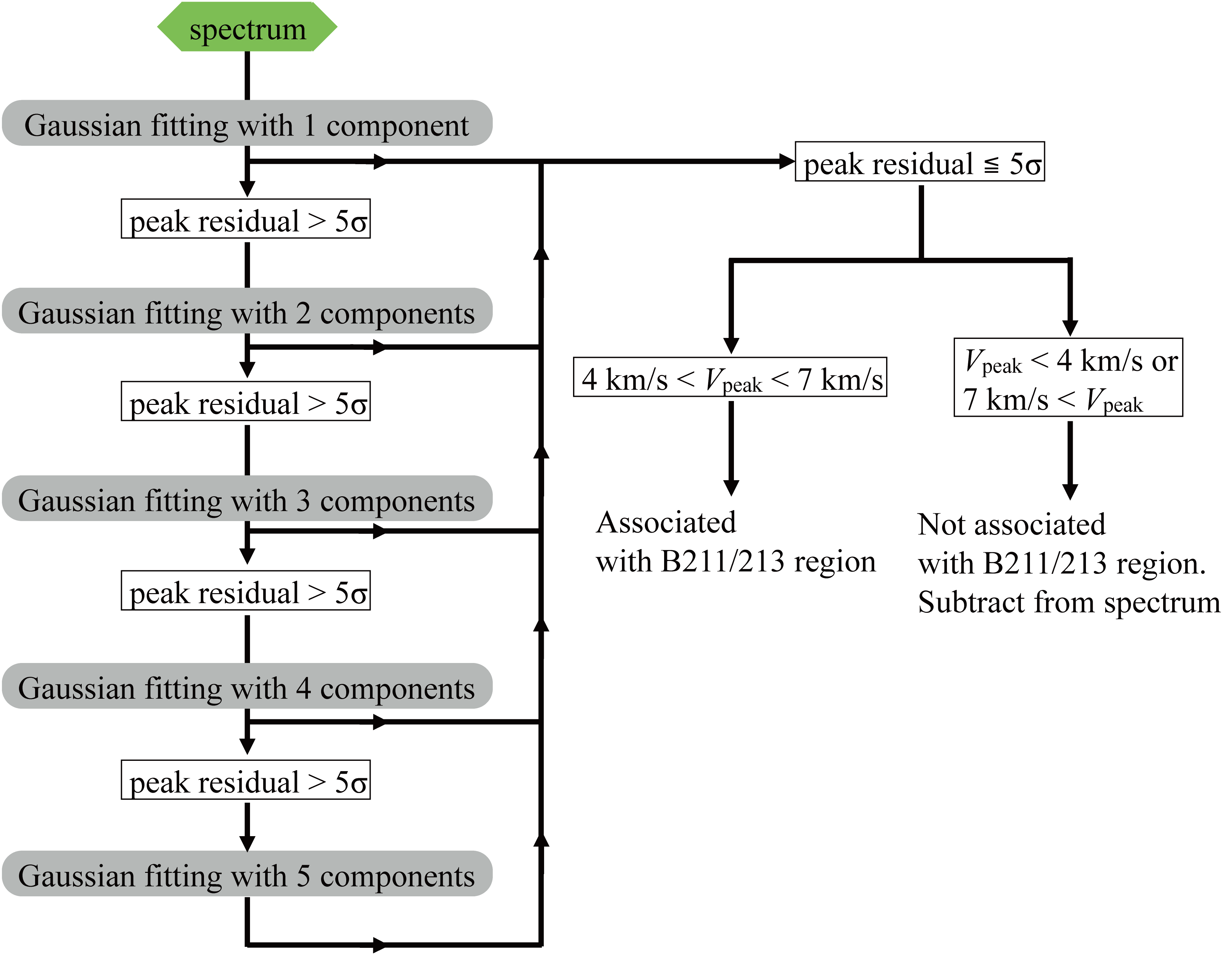}
\caption{Flowchart of our Gaussian fitting procedure with $N$ components. 
Gaussian fitting with $N$=1 component is first applied to each pixel. Where the signal to noise (S/N) ratio of the residual peak intensity is less than 5, we considered that the spectrum at this pixel consist of $N$=1 Gaussian component. When the signal to noise (S/N) ratio of the residual peak intensity is more than 5, we applied Gaussian fitting with ${N+1}$ Gaussian components. This process is repeated up to at most 5 components. 
Then, when the peak velocity is lower than 4.0 km s$^{-1}$ or higher than 7.0 km s$^{-1}$, we consider that the component is not associated with the B211/B213 cloud and subtract it from the data cube.}
\label{fig:fitting}
\end{figure*}

\begin{figure*}
\centering
\includegraphics[width=150mm, angle=0]{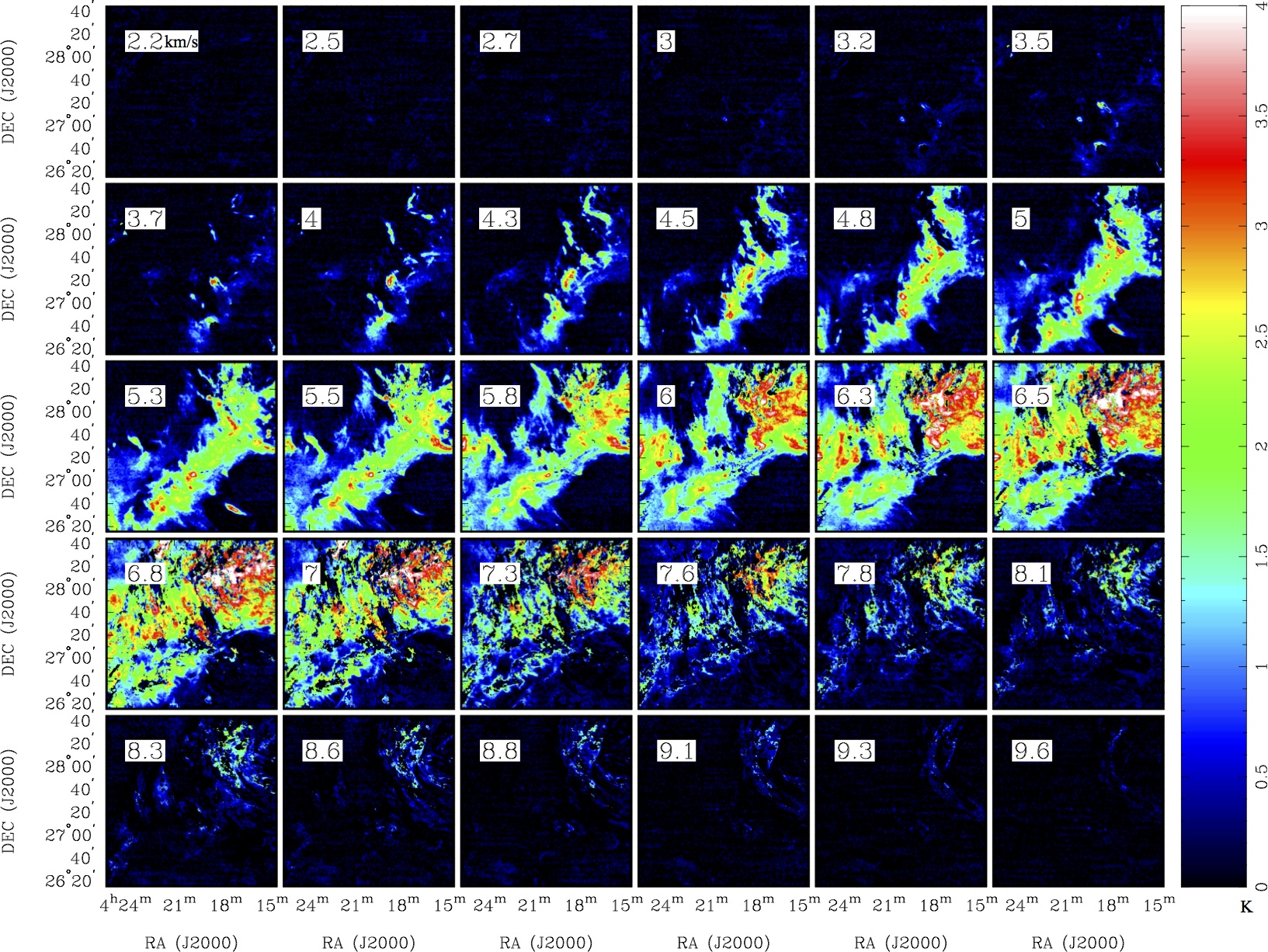}
\includegraphics[width=150mm, angle=0]{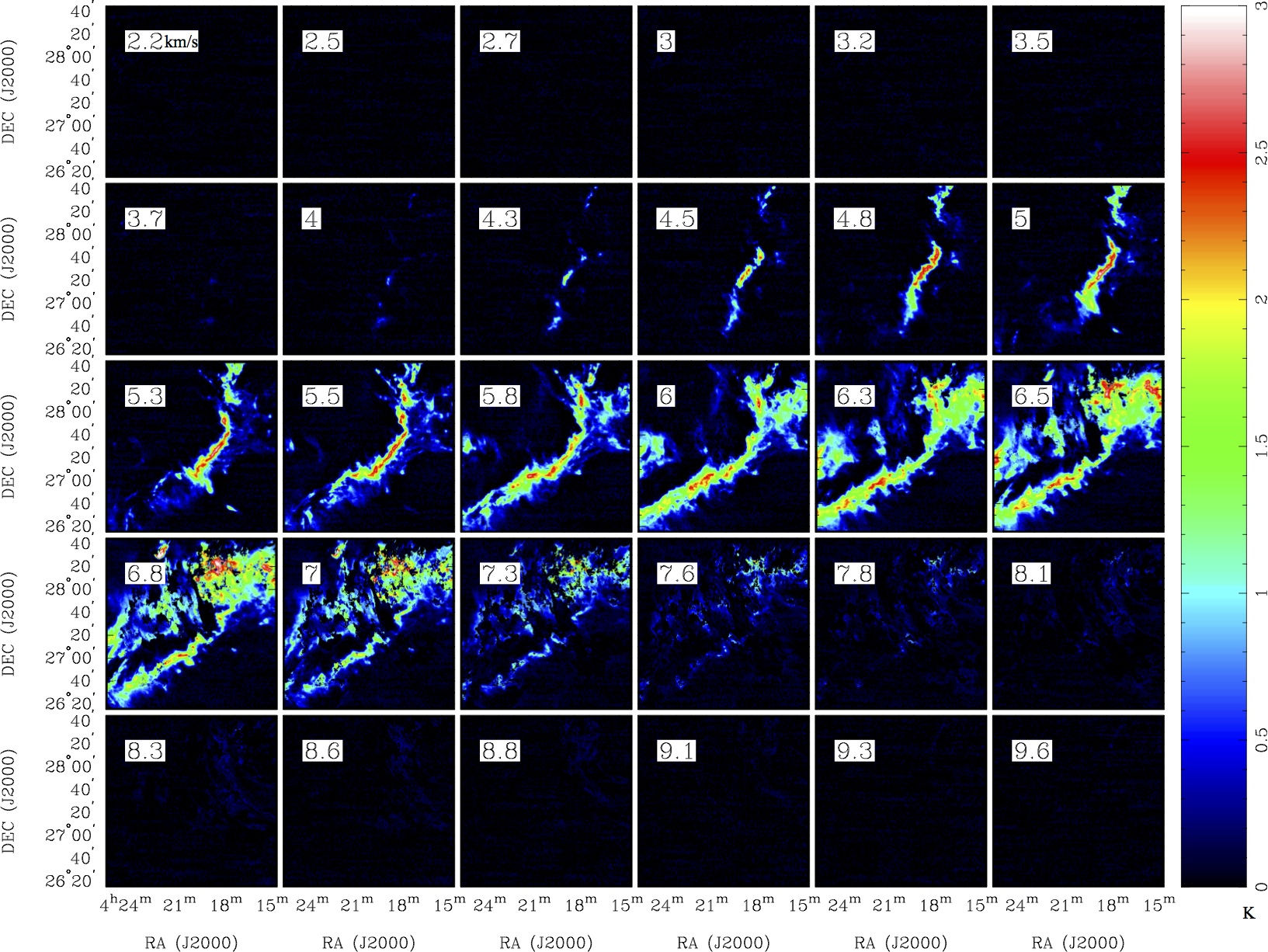}
\caption{$^{12}$CO ($J$=1--0, $top$) and $^{13}$CO ($J$=1--0, $bottom$) velocity channel maps after subtracting the components not associated with the B211/B213 filament. The intensity unit is K. The velocities in km s$^{-1}$ are indicated at the top-left corner of each panel. The velocity width of each channel map is 0.3 km s$^{-1}$.}
\label{fig:channel_subt}
\end{figure*}

\begin{figure*}
\centering
\includegraphics[width=150mm, angle=0]{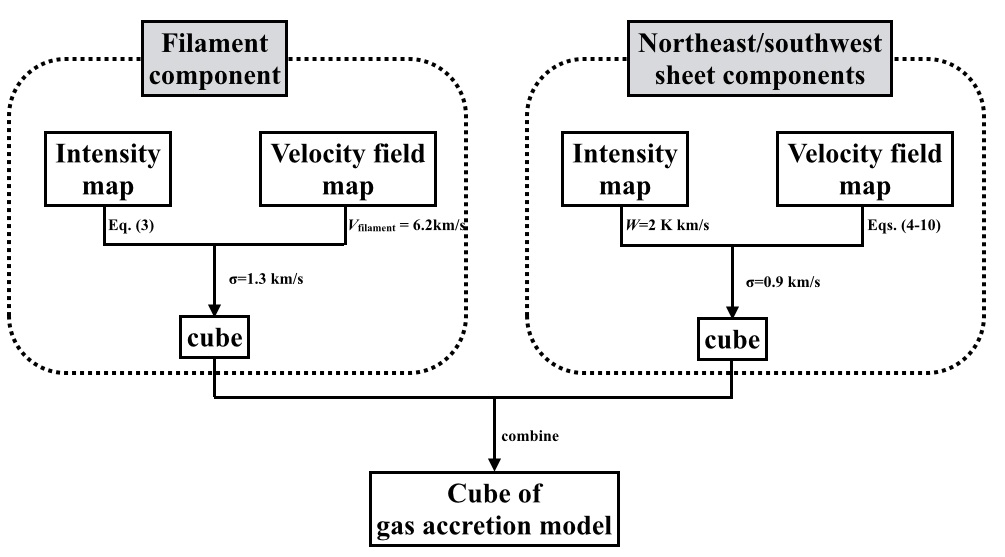}
\caption{Flowchart for the 3-component modeling procedure (see Sect. \ref{section:model}).}
\label{fig:modeling_flow}
\end{figure*}

\begin{figure*}
\centering
\includegraphics[width=180mm, angle=0]{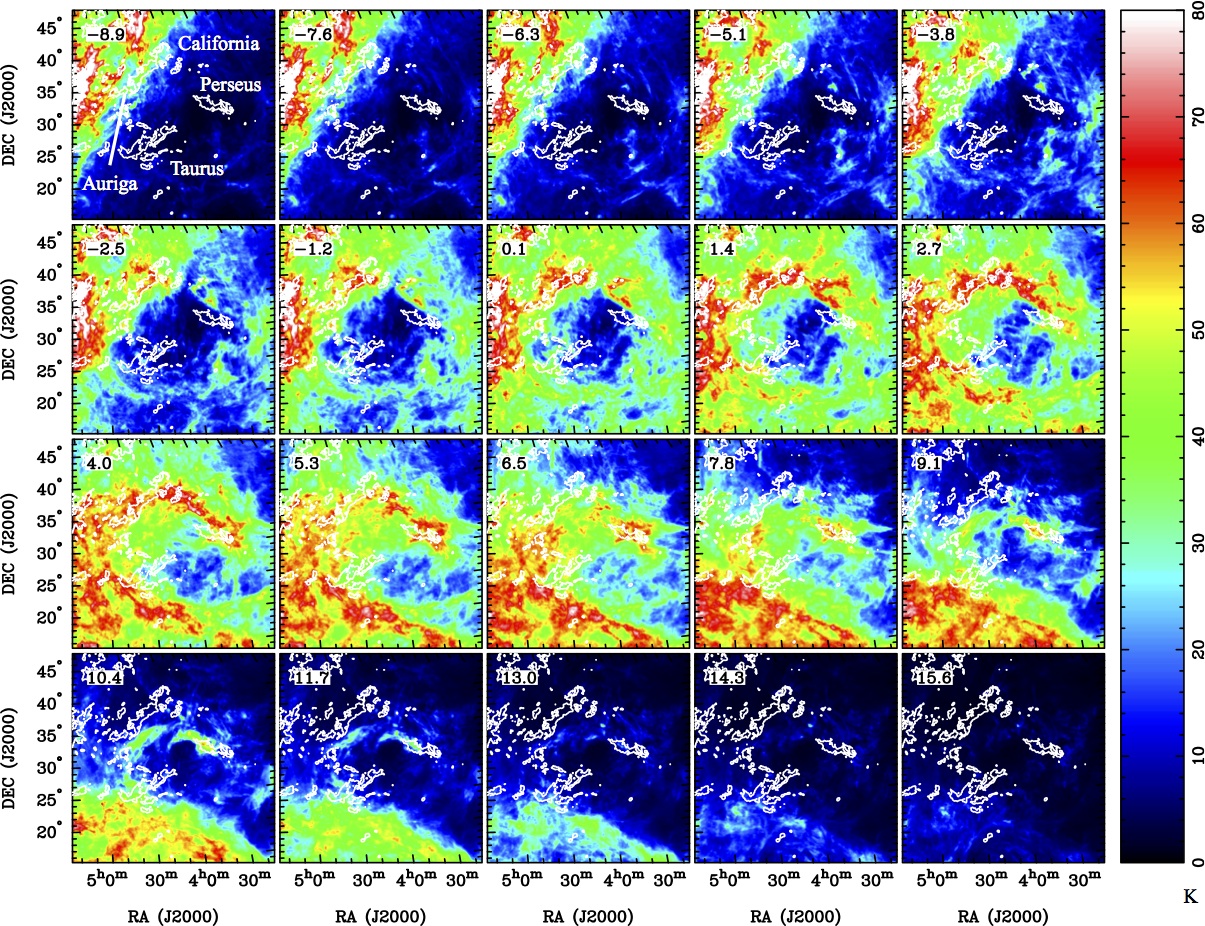}
\caption{HI velocity channel maps. The HI data are from the Effelsberg-Bonn HI Survey (EBHIS) and Galactic All-Sky Survey (GASS) \citep{HI4PI16}. The intensity unit is K. The contour indicates a level of 30 MJy str$^{-1}$ at the $Planck$ 857 GHz emission. The velocities are indicated at the top-left corner of each panel. The velocity width of each channel map is 1.3 km s$^{-1}$.}
\label{fig:channel_hi}
\end{figure*}

\begin{figure}
\centering
\includegraphics[width=80mm, angle=0]{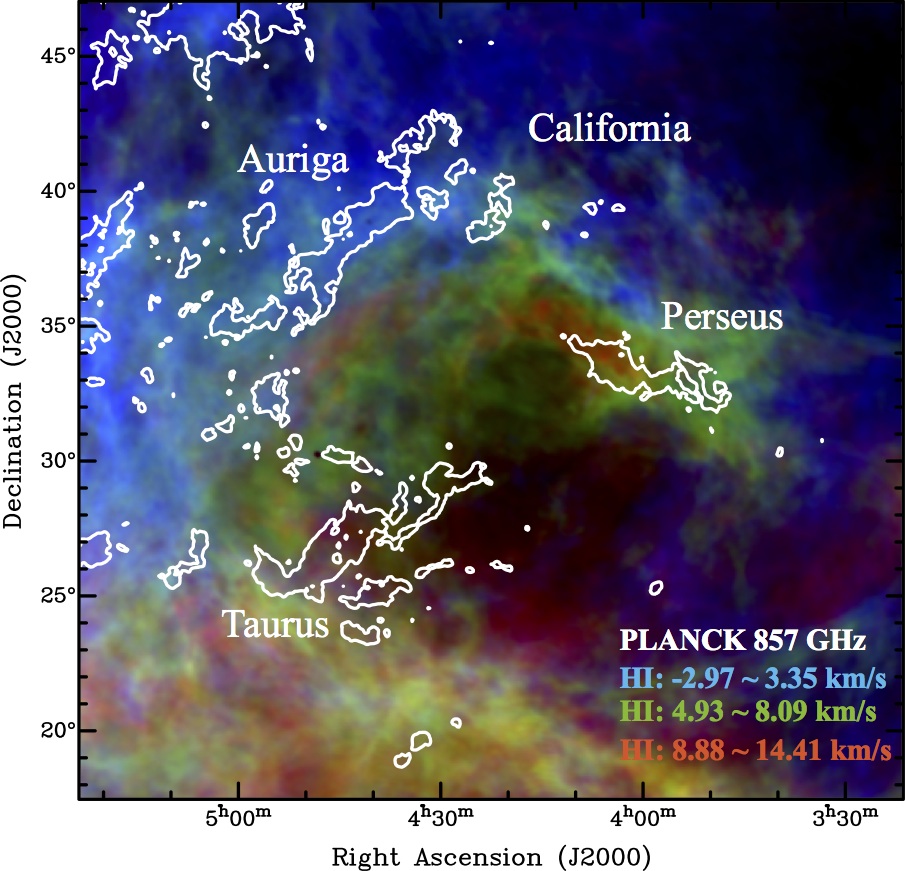}
\caption{Large-scale spatial distribution of HI emission in the Taurus-Auriga-California-Perseus region from \citet{Kalberla17}. Red, green,and blue are the HI components at the velocity of -29.7--3.35, 4.93--8.09, and  8.88--14.41 km s$^{-1}.$ Contour indicates a level of 30 MJy str$^{-1}$ at the $Planck$ 857 GHz emission.}
\label{fig:3color_hi}
\end{figure}

\begin{figure*}
\centering
\includegraphics[width=60mm, angle=0]{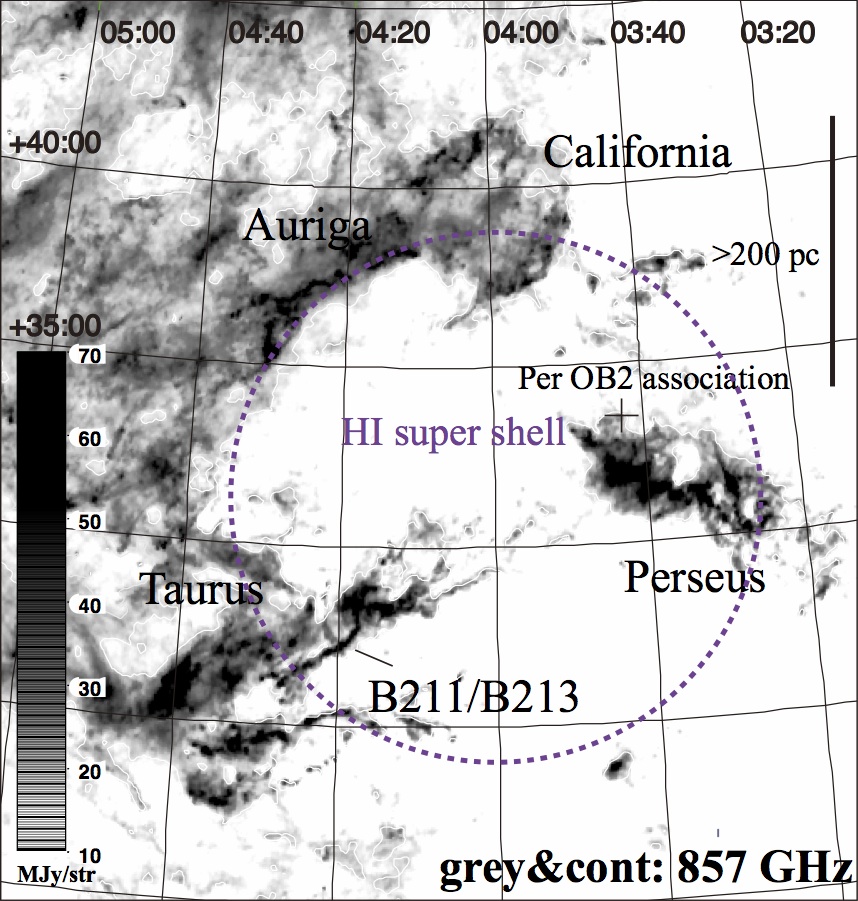}
\includegraphics[width=60mm, angle=0]{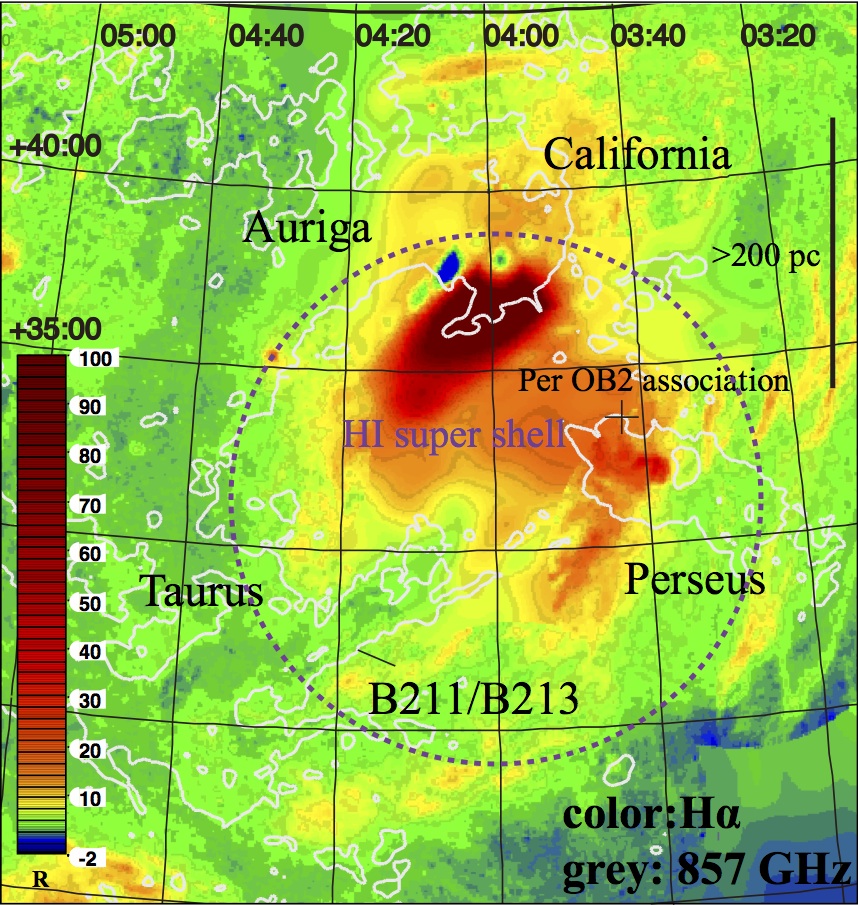}
\caption{Spatial distributions of ($left$) the H${\alpha}$ \citep[color,][]{Finkbeiner03} and ($right$) 857 GHz dust \citep[grey,][]{Planck14} emission. Same as Fig. \ref{fig:halpha}, but the H${\alpha}$ and $Plancl$ 857 GHz maps are displayed separately, for better clarity. Contour indicates a level of 10 MJy str$^{-1}$ at the $Planck$ 857 GHz emission.}
\label{fig:halpha_appendix}
\end{figure*}

\textcolor{black}{
Figure \ref{fig:fitting} is a flowchart of our Gaussian fitting procedure with N components.
Figure \ref{fig:channel_subt} shows the 12CO(1-0) and 13CO(1-0) velocity channel maps after subtracting 
the components not associated with the B211/B213 filament (see Sect. \ref{section:pv_discription}). 
Figure \ref{fig:modeling_flow} is a flowchart for our 3-component modeling procedure described in 
Sect. \ref{section:model}.  
Figure \ref{fig:channel_hi} and Figure \ref{fig:3color_hi} show the large-scale spatial distribution of HI emission 
 in the Taurus-Auriga-California-Perseus region based on the HI data from, \citet[e.g., ][]{Kalberla17}.  
Figure \ref{fig:halpha_appendix} shows the large-scale spatial distributions of the Halpha emission
(Finkbeiner 2003) and 857 GHz dust emission \citep{Planck14} 
(see also Fig. \ref{fig:halpha} and Sect. \ref{Sect:compress}). 
}

\section{Inclinations of the two sheet components in the model}\label{appendix:inclination}

\begin{figure*}
\centering
\includegraphics[width=160mm, angle=0]{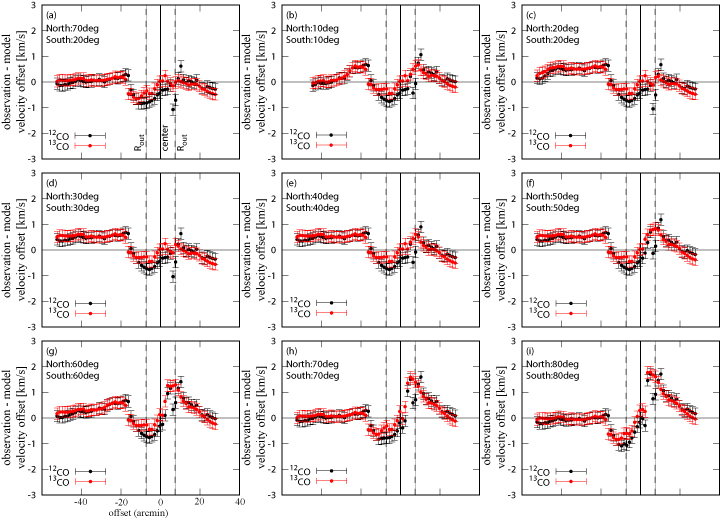}
\caption{Velocity offset as a function of position between the model and $^{12}$CO(1-0)/$^{13}$CO(1-0)  in the $PV$ diagrams. The assumed inclination angle  for the sheet in the model are indicated at the top-left corner of each panel.  The black and red indicate the velocity differences between the model and $^{12}$CO (1--0) data and between the model and $^{13}$CO (1--0) data, respectively.
}
\label{fig:variation}
\end{figure*}

To investigate the effect of the assumed inclinations for the two sheet components in our model, we expected a range of inclinations (10$^\circ$, 20$^\circ$, 30$^\circ$, 40$^\circ$, 50$^\circ$, 60$^\circ$, 70$^\circ$, and 80$^\circ$) and compared, for each inclination, the (peak) velocities predicted by the model with the $^{12}$CO/$^{13}$CO observations. 
The northeastern and southwestern sheet components were examined separately. We assumed the same parameters as listed in Table \ref{Table1}, except for the inclination angles ($\theta_{\rm N/S}$).
Figure \ref{fig:variation} shows the velocity offsets between models and observations. The velocity offset for the southwestern sheet component (offset $>$ 0) increases as the inclination angle increases, while the velocity offset for the northeastern sheet component (offset $<$ 0) increases as the inclination angle decreases. The velocity offset at |offset| $<$ $R_{\rm out}$ tends to be larger than that at |offset| $>$ $R_{\rm out}$. One possible reasons is that the $^{12}$CO (1--0) and $^{13}$CO (1--0) emissions do not trace the inner part of the filament (|offset| $<$ $R_{\rm out}$) since the $^{12}$CO (1--0) and $^{13}$CO (1--0) optical depths are much larger than a unity (see Sect. \ref{optical_depth}). In order to further investigate the velocity fields of the accreting gas, observations in optical thin dense gas tracers such as N$_2$H$^{+}$(1-0) and H$^{13}$CO$^+$(1-0) which trace the filament well \citep[cf.][for H$^{13}$CO$^{+}$ (1--0)]{Shimajiri17} are required. Table \ref{TableA1} summarizes the mean values of the velocity offsets for the northeastern and southwestern sheet components for each model. Inclinations of 70$^{\circ}$ for the northeastern sheet component and 20$^{\circ}$ for the southwestern sheet component provide the minimum velocity offset. We therefore adopted these inclination values in the model presented in Sect. \ref{section:model}.

\begin{table}
\caption{Mean values of the velocity offset between $^{12}$CO (1--0)/$^{13}$CO (1--0) observations and model}             
\label{TableA1}      
\centering                   
\tiny      
\begin{tabular}{cccc}        
\hline\hline                
Inclination$^{\dag}$  & North/South &  line  & Mean$\pm$Stddev$^{\ddag}$ \\   
\hline
\multirow{4}{*}{10$^{\circ}$} & North & $^{12}$CO & 0.48$\pm$ 0.44 \\
                                           & South & $^{12}$CO & 0.29$\pm$ 0.25
 \\
                                           & North & $^{13}$CO & 0.41$\pm$ 0.49
 \\
                                           & South & $^{13}$CO & 0.25$\pm$ 0.19
 \\
\hline
\multirow{4}{*}{20$^{\circ}$} & North & $^{12}$CO & 0.48$\pm$ 0.17 \\
                                           & South & $^{12}$CO & 0.26$\pm$ 0.24
 \\
                                           & North & $^{13}$CO & 0.43$\pm$ 0.17
 \\
                                           & South & $^{13}$CO & 0.17$\pm$ 0.15
 \\
\hline
\multirow{4}{*}{30$^{\circ}$} & North & $^{12}$CO & 0.50$\pm$ 0.14 \\
                                           & South & $^{12}$CO & 0.27$\pm$ 0.24
 \\
                                           & North & $^{13}$CO & 0.45$\pm$ 0.17
 \\
                                           & South & $^{13}$CO & 0.21$\pm$ 0.17
 \\
\hline
\multirow{4}{*}{40$^{\circ}$} & North & $^{12}$CO & 0.50$\pm$ 0.14 \\
                                           & South & $^{12}$CO & 0.26$\pm$ 0.19
 \\
                                           & North & $^{13}$CO & 0.45$\pm$ 0.17
 \\
                                           & South & $^{13}$CO & 0.25$\pm$ 0.19
 \\
\hline
\multirow{4}{*}{50$^{\circ}$} & North & $^{12}$CO & 0.50$\pm$ 0.14 \\
                                           & South & $^{12}$CO & 0.31$\pm$ 0.29
 \\
                                           & North & $^{13}$CO & 0.45$\pm$ 0.17
 \\
                                           & South & $^{13}$CO & 0.37$\pm$ 0.26
 \\
\hline
\multirow{4}{*}{60$^{\circ}$} & North & $^{12}$CO & 0.38$\pm$ 0.23 \\
                                           & South & $^{12}$CO & 0.47$\pm$ 0.41
 \\
                                           & North & $^{13}$CO & 0.33$\pm$ 0.14
 \\
                                           & South & $^{13}$CO & 0.54$\pm$ 0.44
 \\
\hline
\multirow{4}{*}{70$^{\circ}$} & North & $^{12}$CO & 0.26$\pm$ 0.28 \\
                                           & South & $^{12}$CO & 0.62$\pm$ 0.45
 \\
                                           & North & $^{13}$CO & 0.20$\pm$ 0.18
 \\
                                           & South & $^{13}$CO & 0.63$\pm$ 0.52
 \\
\hline
\multirow{4}{*}{80$^{\circ}$} & North & $^{12}$CO & 0.29$\pm$ 0.35 \\
                                           & South & $^{12}$CO & 0.73$\pm$ 0.51
 \\
                                           & North & $^{13}$CO & 0.21$\pm$ 0.29
 \\
                                           & South & $^{13}$CO & 0.74$\pm$ 0.60
 \\
\hline
\hline
\end{tabular}
\tablefoot{(\dag) Assumed inclinations for the northeast and southwest sheet components in the model.}
\tablefoot{(\ddag) Mean and standard deviation of the velocity offset between observations and model.}
\end{table}

\end{document}